\documentclass[english, prb, showpacs, reprint, superscriptaddress]{revtex4-2}
\usepackage[latin9,utf8]{inputenc}
\usepackage{color}
\usepackage{babel}
\usepackage{verbatim}
\usepackage{bm}
\usepackage{amsmath}
\usepackage{amssymb, mathtools}
\usepackage{graphicx}
\usepackage{indentfirst}
\usepackage[unicode=true,pdfusetitle,
bookmarks=true,bookmarksnumbered=false,bookmarksopen=false,
breaklinks=false,pdfborder={0 0 0},pdfborderstyle={},backref=false,colorlinks=true]
{hyperref}
\hypersetup{
	citecolor=blue,urlcolor=black}
\usepackage[normalem]{ulem}

\makeatletter

\usepackage{epstopdf}
\usepackage{amsfonts}
\usepackage{mathrsfs}\usepackage{bm}\usepackage{appendix}
\usepackage{txfonts}
\usepackage{float}
\usepackage{epstopdf}
\usepackage{ulem}

\makeatother
\begin{document}
	\title{An effective curved space-time geometric theory of generic twist angle graphene with application to a rotating bilayer configuration}
	\affiliation{%
		State Key Laboratory of Optoelectronic Materials and Technologies, Guangdong Provincial Key Laboratory of Magnetoelectric Physics and Devices, School of Physics, Sun Yat-Sen
		University, Guangzhou 510275, China\\
		$^{2}$Department of Chemistry and Physics, Augusta University, 1120 15$^{th}$ Street, Augusta, Georgia~30912, USA\\
		$^{3}$Kavli Institute for Theoretical Physics, University of California, Santa Barbara, California 93106, USA\\
		$^{4}$International Quantum Academy, Shenzhen 518048, China.
	}
	\author{Jia-Zheng Ma$^{1}$}
	\author{Trinanjan Datta$^{2,3}$}
	\email{tdatta@augusta.edu}
	\author{Dao-Xin Yao$^{1,4}$}
	\email{yaodaox@mail.sysu.edu.cn}
	\bibliographystyle{plain}
	\begin{abstract}
		\indent\setlength{\parindent}{1em}We propose a new kind of geometric effective theory based on curved space-time single valley Dirac theory with spin connection for twisted bilayer graphene under generic twist angle. This model can reproduce the nearly flat bands with particle-hole symmetry around the first magic angle. The band width is near the former results given by Bistritzer-MacDonald model or density matrix renormalization group. Even more, such geometric formalism allows one to predict the properties of rotating  bilayer graphene which cannot be accessed by former theories. As an example, we investigate the Bott index of a rotating bilayer graphene. We relate this to the two-dimensional Thouless pump with quantized charge pumping during one driving period which could be verified by transport measurement.    \\
	\end{abstract}
	\date{\today}
	\maketitle
	
	
	\section{Introduction}
	\label{sec:introduction}
	\indent\setlength{\parindent}{1em}Twisted bilayer graphene (TBG) has attracted great interest both from a theoretical ~\cite{MacDonald2011pnas} and an experimental perspective~\cite{YuanCao2018Nature,YCao2018Nat}. This system, which is rich in its physical behavior, provides a platform for studying the strongly correlated electronic state~\cite{YuanCao2018Nature}, orbital magnetism~\cite{XiDai2019prx,JianPengLiu2021prb}, superconductivity~\cite{YCao2018Nat} fragile topological phase~\cite{Bernevig2019prl,Ashvin2019prb}, higher-order topological insulator phase~\cite{SungBin2019prl} and higher-order topological superconductivity~\cite{Bernevig2021arxiv}. Furthermore, heterostructure based on TBG may have potential application in superconducting devices~\cite{YCao2021Naturenano} and quantum computation~\cite{Alicea2022prb}. While several theories have been proposed to explain the occurrence of electronic phases in TBG, amongst them, the Bistritzer--MacDonald (BM) model and its descendants ~\cite{Guinea2012prl,Ashvin2019prl,NiuQian2021prl} have been successful in explaining electron localization near the magic angle. The BM model is valid only for small to moderate twist angles. A real space effective field theory formalism of TBG, limited to small deformation gradient has also been constructed~\cite{Balents2019SciPost,Olle2022arxiv}. Furthermore, the curved space quantum field theory (QFT) formalism has been utilized to explain the magic continuum within the context of staggered flux twisted bilayer square lattice~\cite{CenKeXu2021prb}. However, a real space field theory formalism for generic twist angle in a
	moir\'e system is still missing. Thus, there is a need for a generic twist angle theory for TBG and twisted bilayer bravais lattice~\cite{Ashvin2019prr}.
	\\
	\indent\setlength{\parindent}{1em}
	The need for the generic twist angle theory becomes even more apparent if we consider an out of equilibrium system. In this context, optical Floquet engineering~\cite{Fiete2020prb,LedeXian2019prr} and the Thouless pump~\cite{DiXiao2020prb} physics of TBG have been studied. Since a theory based on commensurate approximation~\cite{Castro2012prb} is not enough to capture the incommensurate nature for any twisted angle, there is a need to develop a geometric theory for generic twisted angle. Such a formalism can be applied to non-adiabatic rotation, including a structurally rotating TBG which we call the rotating bilayer graphene (RBG). \\
	\indent\setlength{\parindent}{1em}
	We obtain the deformation field under an arbitrary twist angle and the energy bands at the first magic angle based on curved space-time Dirac action for a non-interacting TBG and RBG model. We utilize a geometric method to study the physics of TBG. This method is in principle valid for arbitrary twist angle, which is mainly inspired by considering the twist as a kind of deformation~\cite{Balents2019SciPost} and attributing the approximate zero energy flat band to the effective SU(2) gauge field (pseudo magnetic vector potential) in TBG~\cite{Guinea2012prl,Jian-PengLiuXi-Dai2019prb,QuanShengWu2021prl,CenKeXu2021prb,YaoWang2020prl,Shaffique2019prb,Leonardo2021prb,Leonardo2022prb}. Unlike the BM model, our geometric model can be non-Hermitian and break the PT symmetry. Thus, the $\bf K_{2}$ point in the moir\'e brillouin zone will be gapped out. Such a geometric theory can be naturally generalized to the case of a RBG system, which cannot be modeled by previous theoretical formulations. By calculating the Bott index, we obtain the quantized charge pumping in RBG.
	Within our effective geometric theory, the emergent SU(2) gauge field in RBG will generate a spin connection which will mix with the Aharonov-Anandan connection in RBG. The total connection will be the summation of these two components with Aharonov-Anandan connection and may give rise to new topological phases as well as new type of Floquet engineering in TBG (compared to Floquet engineering based on optical driving~\cite{Fiete2020prb}). For simplicity, we consider a non-interacting model as a first attempt to generalize the TBG geometric formalism. The key point is to recover the flat bands and generalize to a RBG configuration.
	\\ 
	\indent\setlength{\parindent}{1em}This article is organized as follows. In Sec.~\ref{sec:TBG} we briefly review the theoretical development of TBG. In Sec.~\ref{sec:vierbein} we show how to obtain the deformation field under arbitrary twist angle and the corresponding vierbein. In Sec.~\ref{sec:Diraceq} we solve the curved space Dirac equation at the first magic angle as a benchmark. We also compute the band structure for the $30^{\circ}$ quasicrystal TBG, showing that our model is applicable to an incommensurate system. In Sec.~\ref{sec:RBG} we introduce the RBG model and compute the Bott index as an indicator of topological charge pumping. In Sec.~\ref{sec:conclude} we provide our conclusions and discussions. 
	In Appendix~\ref{sec:nonhermiticity} we show the non-Hermiticity of curved space Dirac equation discretization. While in Appendix~\ref{sec:detail} we provide details on the commutations of mixed second order derivative of the deformation field in RBG. Finally, in Appendix~\ref{app:appc} we discuss the meaning of imaginary Fermi velocity and interlayer coupling.
	\section{A brief review of twisted bilayer graphene}\label{sec:TBG}
	In this section we provide a brief review of the TBG focusing exclusively on theoretical and numerical developments. The first tight binding (non-interacting) model was proposed in Ref.~\cite{MacDonald2011pnas}. For small twisted angle, only three dominant momentum transfers were considered. The interlayer moir\'e modulated coupling was projected to momentum space and considered as a perturbation. Next, Guinea \emph{et al.} showed that the moir\'e coupling in TBG can be regarded as a SU(2) gauge field which is responsible for band flattening in TBG~\cite{Guinea2012prl}. Subsequently, all the magic angles and the analytic ground state wavefunction around magic angles were derived ~\cite{Ashvin2019prl}. Furthermore, the magic angle was obtained by combining Wenzel-Kramers-Brillouin approximation (WKB) and asymptotic Airy function solution with single value condition~\cite{NiuQian2021prl}. Recently, several phenomenological many body interacting models for magic angle TBG (MATBG) considering long range Coulomb interaction have also been proposed. Ref~\cite{Liang-Fu2018prx} considers TBG as an extended Hubbard model on triangular superlattice. For MATBG, Jian Kang~\emph{et~al.} have established a U(4) many body model~\cite{JianKang2019prl}. Bernervig~\emph{et~al.} generalized the Kang-Vafek model with more exotic excitations via quantum geometric method~\cite{Bernevig2021prb}. Furthermore, several numerveical methods have been developed to solve the TBG many body Hamiltonian including exact diagonalization (ED)~\cite{MacDonald2021prl}, determinant quantum Monte Carlo (DQMC)~\cite{ZiYangMeng2021prx,pay2021cpl,JongYeonLee2022prx}, dynamical mean field theory (DMFT)~\cite{vahedi2021SciPostPhys}, density matrix renormalization group (DMRG)~\cite{Zaletel2020prb,Vafek2020prb,WeiLi2021natcomm}. Furthermore, a real space formalism has been developed regarding the twist as a special deformation~\cite{Balents2019SciPost}. This theory has the advantage of characterizing the effect of relaxation. Its descendant has been used in twisted bilayer staggered flux square lattice~\cite{CenKeXu2021prb} (a model of twisted bilayer spin liquid) and explains the magic continuum. In addition to the above, Ref.~\cite{Grandi2021JHEP} has proposed holographic duality construction of flat band and revealed the presence of nematic order. Furthermore, Ref.~\cite{gaa2021prb} based on fracton-elasticity duality has explained TBG quasicrystal elasticity. Ref.~\cite{BoYang2021prl} implemented a geometric method based on Fubini-Study metric approach and deduced the Landau zero energy flat band with interaction in TBG. There has also been a calculation based on the vierbein formalism which demonstrates that there exists emergent moir\'e gravity in strained TBG~\cite{Galitski2022prrletter}. Ref.~\cite{Kumira2016ptep} connection the BM model with the data in string theory.  
	\section{The vierbein formalism for TBG and the deformation field}\label{sec:vierbein}
	Possible sources of deformation in continuous media include rigid twist, relaxation, and slide. We consider only the effects of a rigid twist which is defined by $\bf u=\phi\times r$, where $\bf u$ is the deformation field, $\bf \phi$ is the twist angle, and $\bf r$ is the position. In our model we ignore relaxation and slide. Neglecting these additional forms of deformations amount to ignoring certain physical features. Since the Wannier centers are located in the AA regions of the TBG~\cite{Liang-Fu2018prx}, this implies that the AA region will have a larger Coulomb repulsion. So the in-plane relaxation will shrink the AA region and expand the AB region. Similarly, the out-of-plane relaxation will enlarge the AA region interlayer distance while decrease the AB region counterpart. Additionally, the slide is the relative translation between two layers~\cite{DiXiao2020prb}. To avoid the above complexities, we construct the vierbein formalism of the deformation field in TBG in the absence of lattice relaxation and torsion. 
	
\begin{figure}[t]
		\includegraphics[width=1\columnwidth]{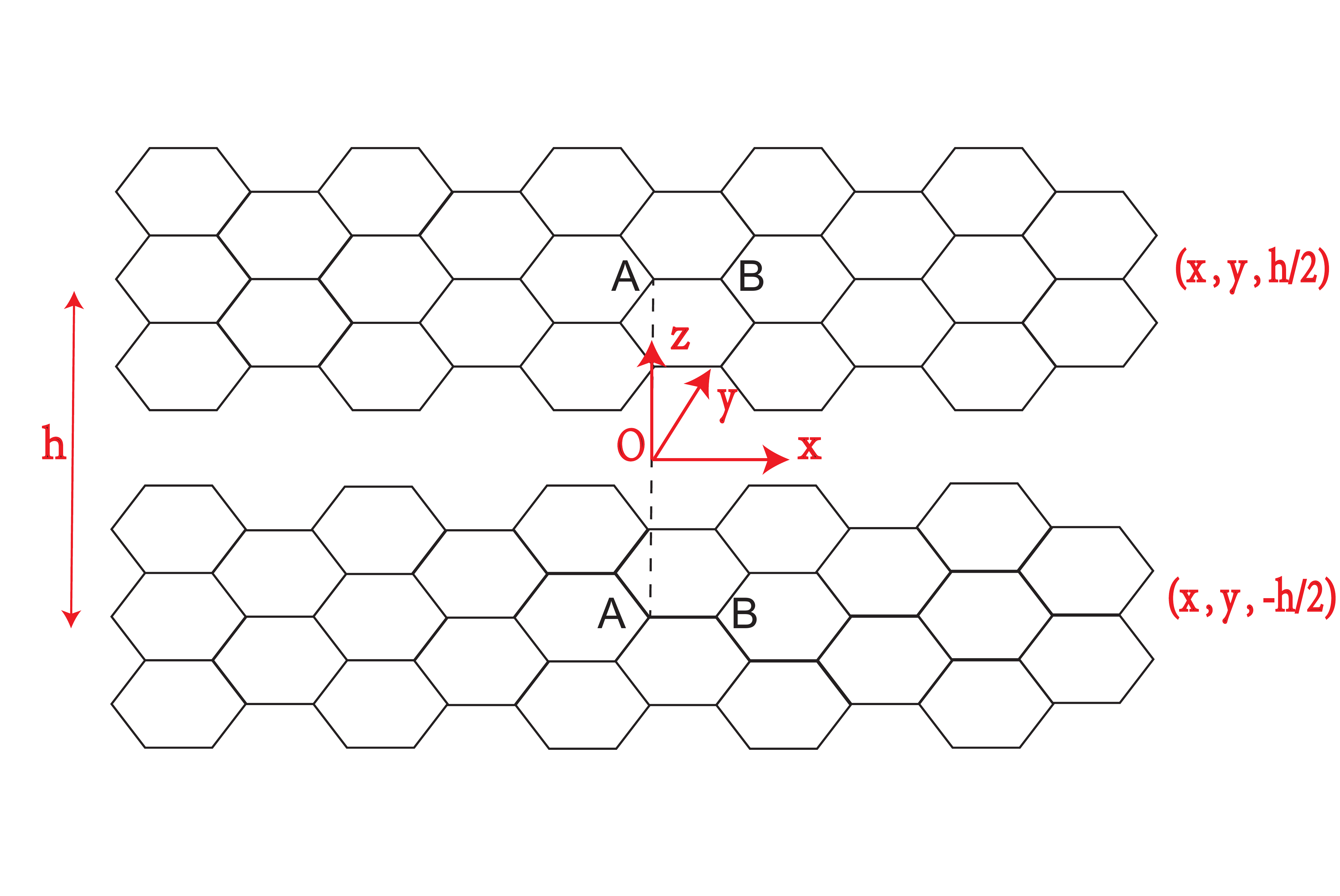}
		\caption{The coordinate notation sketch. The origin is the cross point of rotation axis and middle plane $z=0$. Two layers lay in $z=\pm h/2$ respectively. The rotation axis goes through the sites of A sublattice from  each layer for untwisted AA stacking bilayer graphene.}
		\label{fig:coordinate}
	\end{figure}

	The necessity of introducing vierbein and curved space-time is to faithfully describe the geometric response of Dirac fermion in real space. Compared to the BM model, the vierbein formalism is applicable for generic twisted angle which may have no translation symmetry. When compared to the real space model formulation~\cite{Balents2019SciPost}, the vierbein formalism can handle the system even with singular deformation field $|\partial_{i}\textbf{u}|>>1$. Thus, the curved space-time view can characterize the complicated deformation field in TBG concisely. \\
	\indent\setlength{\parindent}{1em}
	To set up the vierbein formalism, we need to choose an axis of rotation. As shown in Fig.~\ref{fig:coordinate} the rotation axis is at the middle plane $z=0$ of TBG. We choose this as the origin. The two layers are located at $z=\pm\frac{h}{2}$ respectively. The rotation axis goes through the sites of A sublattice from  each layer for untwisted AA stacking bilayer graphene. The sublattice notation for untwisted AA stacking bilayer graphene is indicated by Fig.~\ref{fig:coordinate}.
	\begin{figure*}[htp!]
		\includegraphics[width=2\columnwidth]{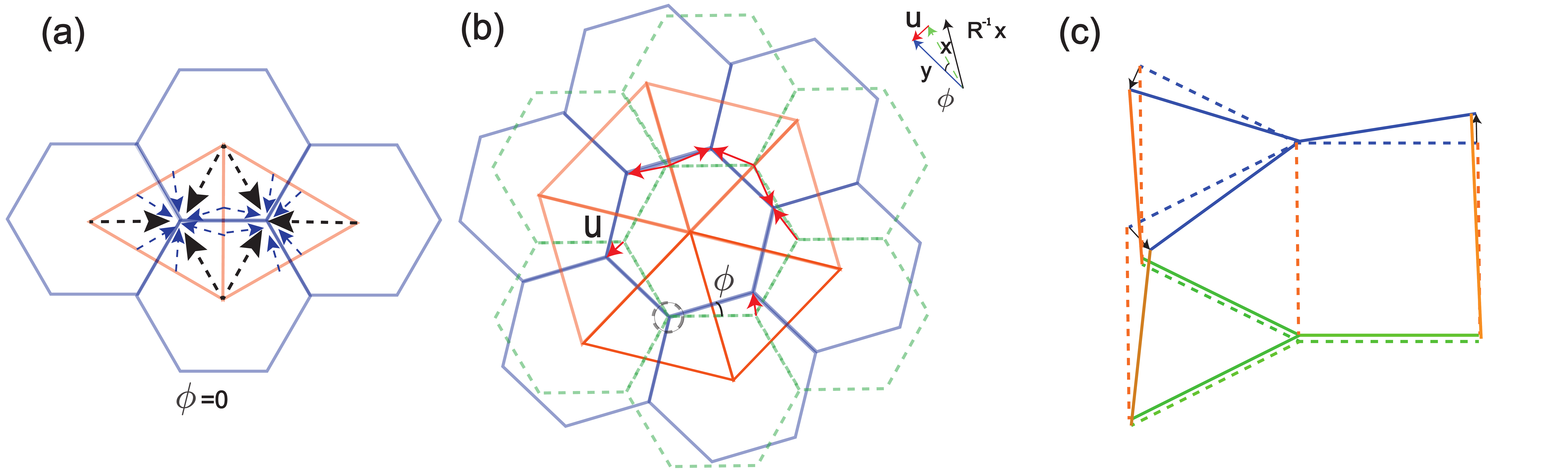}
		\caption{Bilayer configuration and twist geometries. (a) Untwisted AA-stacking bilayer graphene where the two layers overlap on top of each other. The orange triangles are the deformation domain wall. The dashed arrows show the deformation vector field. (b) Twisted bilayer graphene with twist angle $\phi$. The solid blue honeycomb lattice denotes the top layer (twisted one). The dashed green counterpart is the bottom layer (fixed one). The orange triangles correspond to the twisted deformation domain wall. The deformation vector for a given site can be obtained by connecting the site in the untwisted lattice to the nearest center of the triangle domain wall in the twisted lattice. The minifigure in (b) shows the relation between vectors $\bf{x},~\bf{u},~\bf{y}$, and $R^{-1}\bf{x}$ at a given point. (c) A 3d sketch of the twist in TBG. The dashed lines represent the untwisted lattice. The curved space induced by the rigid twist is given by the solid line. The vierbein is induced by its deviation from the dashed line (the untwisted bilayer) according Eq.~\eqref{eq:vierbein}. The black arrows indicate the deformation field. The bottom layer is always fixed. Only the nearest four sites from the rotation axis are shown for each layer.}
		\label{fig:twistedmesh}
	\end{figure*}
	Then we consider the homogeneous twist (deformation) along the $z$ axis. The interlayer distance is $h$. Based on the theory of elasticity, we define the deformation field $\bf{u}$ at a given point as the oriented vector connecting the point before deformation and the counterpart after deformation. Thus the transformation of the deformation field under rotation is given by the following
	\begin{subequations}
		\begin{equation}
		\left(\begin{array}{cccc}u_{x}(\bf{r})\\u_{y}(\bf{r}) \end{array}\right)=R\left( \begin{array}{cccc}u_{x}(R^{-1}\bf{r})\\u_{y}(R^{-1}\bf{r}) \end{array}\right),~
		R= \left( \begin{array}{cccc}\cos\phi&-\sin\phi\\ \sin\phi&\cos\phi\end{array}\right),
		\end{equation}
		\begin{equation}
		R^{-1}\bf{r}=\left( \begin{array}{cccc}
		x\cos\phi+y\sin\phi\\
		-x\sin\phi+y\cos\phi
		\end{array}\right). \label{eq:transformation}
		\end{equation}
	\end{subequations}
	where $\phi$ is the twist angle. Note that, Eq.~\eqref{eq:transformation} is just the transformation rule of a planar vector field. So after the deformation, the coordinates shift to $y^{i}(\bf{r})$ with $\bf{r}=\{x, y, z\}$, as shown in Fig.~\ref{fig:twistedmesh}. The component of deformation vector is denoted by $i$. One can by definition get the vierbein and metric of the effective curved space. For simplicity, we choose the gauge for the vierbein and let it equal to the Jacobian transformation between the curved coordinate and flat counterpart~\cite{PengYe2019prb,ZMH2019prb,Volovik2019prr,Yi-ZhuangYou2016prb,Katanaev2021arxiv}. So the curved space coordinate and vierbein can be expressed as follows
	\begin{eqnarray}
	y^{i}(\bf{r})&=&x^{i}+u^{i}(\bf{r}),\nonumber\\
	e^{\mu}_{a}&=&\frac{\partial y^{\mu}}{\partial x^{a}}=\delta^{\mu}_{a}
	+\partial_{a}u^{\mu},\quad\xi_{\mu}^{a}=(e^{\mu}_{a})^{-1},\label{eq:vierbein}	
	\end{eqnarray}	
	\indent\setlength{\parindent}{1em}The Greek alphabets $\mu,\nu\cdots=0,1,2,3$ denote components of the curved space-time coordinate, while the Latin alphabets $a,b\cdots=0,1,2,3$ denote the flat space-time counterpart or the internal indices of pseudo-spin. 0 denotes time component and 1,2,3 denotes $x,~y,~z$ spatial components respectively. We assume there is a tight binding picture and the lattice is a rigid structure. Although previous formulations have suggested that lattice relaxation helps to isolate the flat band from other higher bands~\cite{Liang-Fu2018prx,Jian-PengLiuXi-Dai2019prb,HongMingWeng2022prb,ochoa2022prl}, for simplicity we do not consider lattice relaxation here. Note, since lattice relaxation can significantly modulate electron-phonon interaction and introduce strain in TBG, there may be non-vanishing torsion under our geometric theory. The layer may be corrugated and thus no longer form a plane. This can lead to added complexity. However, one can improve on this by introducing an out of plane deformation $u_{z}$~\cite{JianPengLiu2021prb} to capture the physics of corrugated TBG. Thus, we ignore lattice relaxation for simplicity.
	\\
	\indent\setlength{\parindent}{1em}
	To apply the vierbein formalism, one should define the deformation field in a discrete lattice. In contrast to continuous media, we define the deformation for a given site as the vector connecting the nearest site after deformation and the original site prior to the deformation. Furthermore, as explained previously we consider this rotation be rigid. The bottom layer is always fixed, which implies that the deformation is zero at each site. However the derivative with $z$ for the deformation in the bottom layer is not zero in general. For the top layer, the deformation is given by connecting the nearest twisted sites and points before the rotation. As illustrated in the Fig.~\ref{fig:twistedmesh}(c), the rotation axis is the center of the AA stacking region. The dashed line corresponds to  the bilayer before the twisting (AA stacking). The solid line represents the twisted bilayer lattice. The black arrows show the deformation of each site in the top layer. For the small radius and the small twisted angle, the deformation is given by $\vec{u}=\vec{\phi}\times\vec{r}$. Notice that for certain twist angles and positions, there may be more than one ``nearest" site. Under such a scenario there is a deformation singularity. We call the position (line) with two ``nearest" sites as a domain wall.
	
	To illustrate the deformation singularity~\cite{davis2021arxiv} at the domain walls of the twisted bilayer lattice, one can draw the Wigner-Seitz cell (dual triangular mesh) of the honeycomb lattice. For the points on the untwisted triangular mesh, the deformation field will be multi-valued. Particularly, points on edges of domain walls have two values while points on vertices of domain walls have six values, see Fig.~\ref{fig:twistedmesh}. Then one can write the deformation field for a given point by connecting it with the nearest center of the twisted triangular mesh in Fig.~\ref{fig:twistedmesh} to obtain
	\begin{figure*}[htp!]
		\includegraphics[width=2\columnwidth]{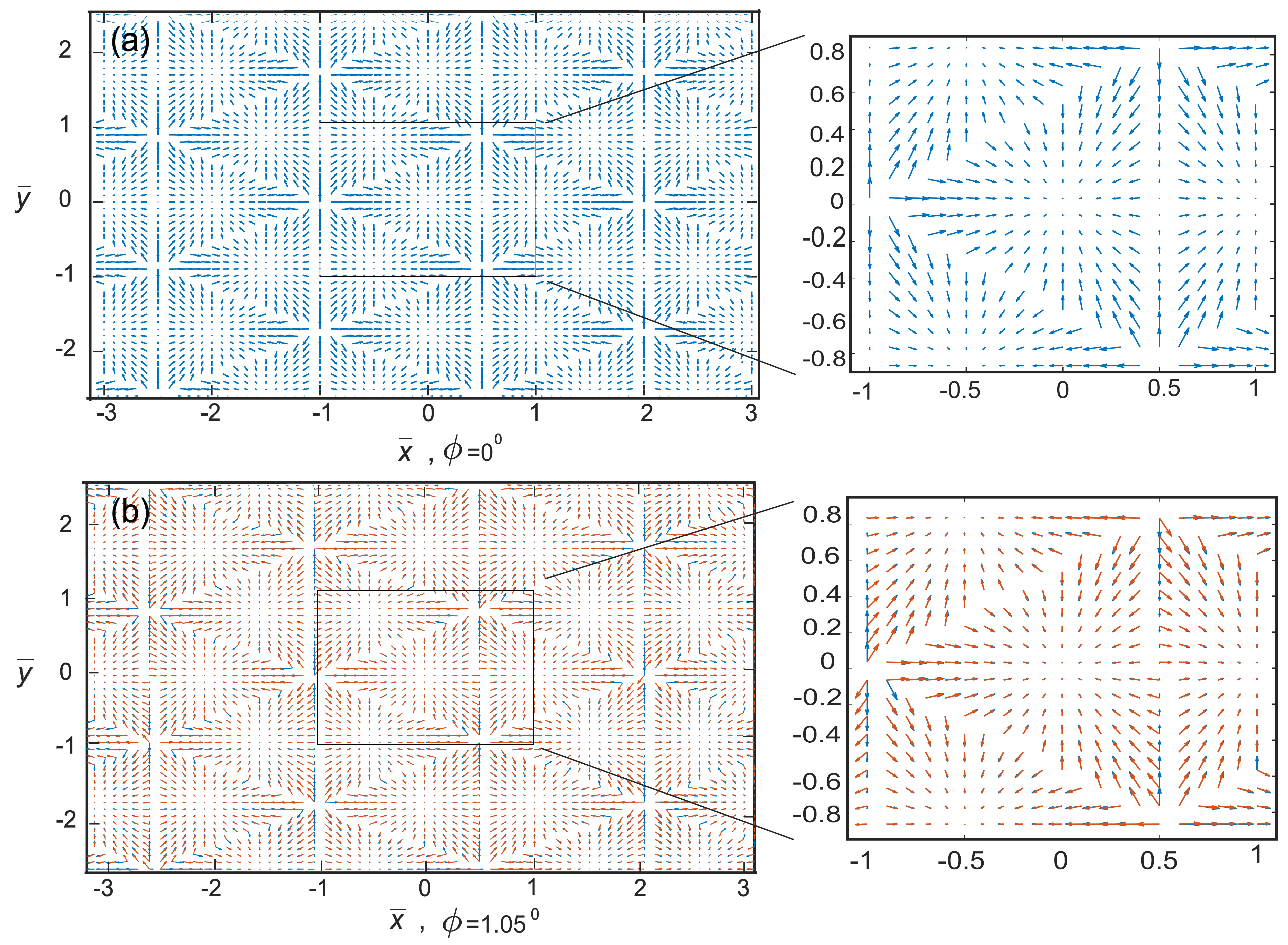}
		\caption{Plot of the top layer deformation field in the whole plane for AA stacking bilayer graphene according to Eqs.~\eqref{eq:deformationvectorA}~--~Eq.~\eqref{eq:deformationvector} for magic angle $\phi$. The arrows show the local deformation vector $\bf{u}$ in Eq.~\eqref{eq:vierbein}. One can observe that the domain walls of deformation form a triangular mesh. Deformation field in (a) zero twisted angle and (b) the first magic angle given by Eqs.~\eqref{eq:deformationvectorA}~--~Eq.~\eqref{eq:deformationvector}. Note, $\bar{x}=x/1.24~\text{\r{A}},~\bar{y}=y/1.24~\text{\r{A}}$ are the dimensionless length in units of the nearest-neighbor distance between carbon atoms in graphene monolayer without relaxation. The inset figure in (a) shows that the deformation field for the two layers coincide. While the inset in (b), shows that the deformation field difference between the two layers is only obvious near the domain walls. The blue arrows represent the deformation of the fixed layer and its red counterpart indicates the deformation of the twisted layer. One can see that there is a kink structure near the domain wall for the red arrows in the zoomed region of (b).}
		\label{fig:wholeplanedeformation}
	\end{figure*}
	\begin{subequations}
		\begin{equation}
		\begin{split}
		u_{x}(\textbf{r}')=\Bigg[\left(\frac{3}{2}(p+q)-x_{1}\right)\vartheta\left(\frac{3}{2}(p+q)+\frac{1}{2}-x_{1}\right)\\ \vartheta\left((x_{1}-\sqrt{3}x_{2})-(3p-1)\right) \vartheta\left((x_{1}+\sqrt{3}x_{2})-(3q-1)\right)+\\\left(\frac{3}{2}(p+q)+1-x_{1}\right)
		\vartheta\left(x_{1}-(\frac{3}{2}(p+q)+\frac{1}{2})\right)\\ \vartheta\left((3p+2)-(x_{1}-\sqrt{3}x_{2})\right)\vartheta\left((3q+2)-(x_{1}+\sqrt{3}x_{2})\right)\Bigg]\\
		\times\left(\frac{z}{h}+\frac{1}{2}\right)\vartheta\left(z+\frac{h}{2}\right)\vartheta\left(\frac{h}{2}-z\right).
		\label{eq:deformationvectorA}
		\end{split}
		\end{equation}
		\begin{equation}
		\begin{split}
		u_{y}(\textbf{r}')=\Bigg[\left(\frac{\sqrt{3}}{2}(q-p)-x_{2}\right)\times\vartheta\left((x_{1}-\sqrt{3}x_{2})-(3p-1)\right)\\ \vartheta\left((x_{1}+\sqrt{3}x_{2})-(3q-1)\right)\vartheta\left((3p+2)-(x_{1}-\sqrt{3}x_{2})\right)\\ \vartheta\left((3q+2)-(x_{1}+\sqrt{3}x_{2})\right)\Bigg]
		\left(\frac{z}{h}+\frac{1}{2}\right)\vartheta\left(z+\frac{h}{2}\right)\vartheta\left(\frac{h}{2}-z\right).
		\label{eq:deformationvectorB}
		\end{split}
		\end{equation}
		\begin{equation}
		\begin{split}
		p=\text{floor}\left(\frac{x_{1}-\sqrt{3}x_{2}+1}{3}\right),\quad
		q=\text{floor}\left(\frac{x_{1}+\sqrt{3}x_{2}+1}{3}\right),\\
		x_{1}=x\cos\phi+y\sin\phi,\quad x_{2}=-x\sin\phi+y\cos\phi,\label{eq:deformationvector}
		\end{split}
		\end{equation}
	\end{subequations}
	where $\vartheta(x)$ is the Heaviside function which describes the location of the triangular mesh (domain walls of the deformation field) and $\text{floor}(x)$ means the least integer function. By examination one can see that $(\partial_{x}\partial_{y}-\partial_{y}\partial_{x})\textbf{u}\neq 0$, which implies a non-vanishing spin connection in the effective curved space. The coordinates are related by $\textbf{r}'=(x_{1},x_{2},z)^{T}=R^{-1}\textbf{r}$ and $\textbf{r}=(x,y,z)^{T}$. The deformation field $\textbf{u}(\textbf{r})$ in the laboratory frame can be obtained using Eq.~\eqref{eq:transformation}. As a check, we can set $\phi=0$ in Eqs.~\eqref{eq:deformationvectorA}~--~\eqref{eq:deformationvector}. With this substitution one recovers the deformation field for the untwisted bilayer graphene. In Fig.~\ref{fig:twistedmesh}(a), the deformation vectors in the two neighboring triangular cells have been drawn only. \\
	\indent\setlength{\parindent}{1em}A detailed distribution of the deformation field is presented in Fig.~\ref{fig:wholeplanedeformation}. The field for each site is shown by an arrow which connects the site in an untwisted lattice and the center of the nearest triangular domain wall with its twisted counterpart. For the untwisted site located on the triangular domain wall mesh, it will have a singular deformation. We observe that the deformation domain wall forms a triangular mesh in the whole plane. When crossing the domain wall, the deformation field component which is perpendicular to the domain wall changes direction while the component parallel to domain wall remains unchanged. 

	\section{Dirac equation with spin connection}
	\label{sec:Diraceq}
	\begin{figure*}[htp!]
		\includegraphics[width=2\columnwidth]{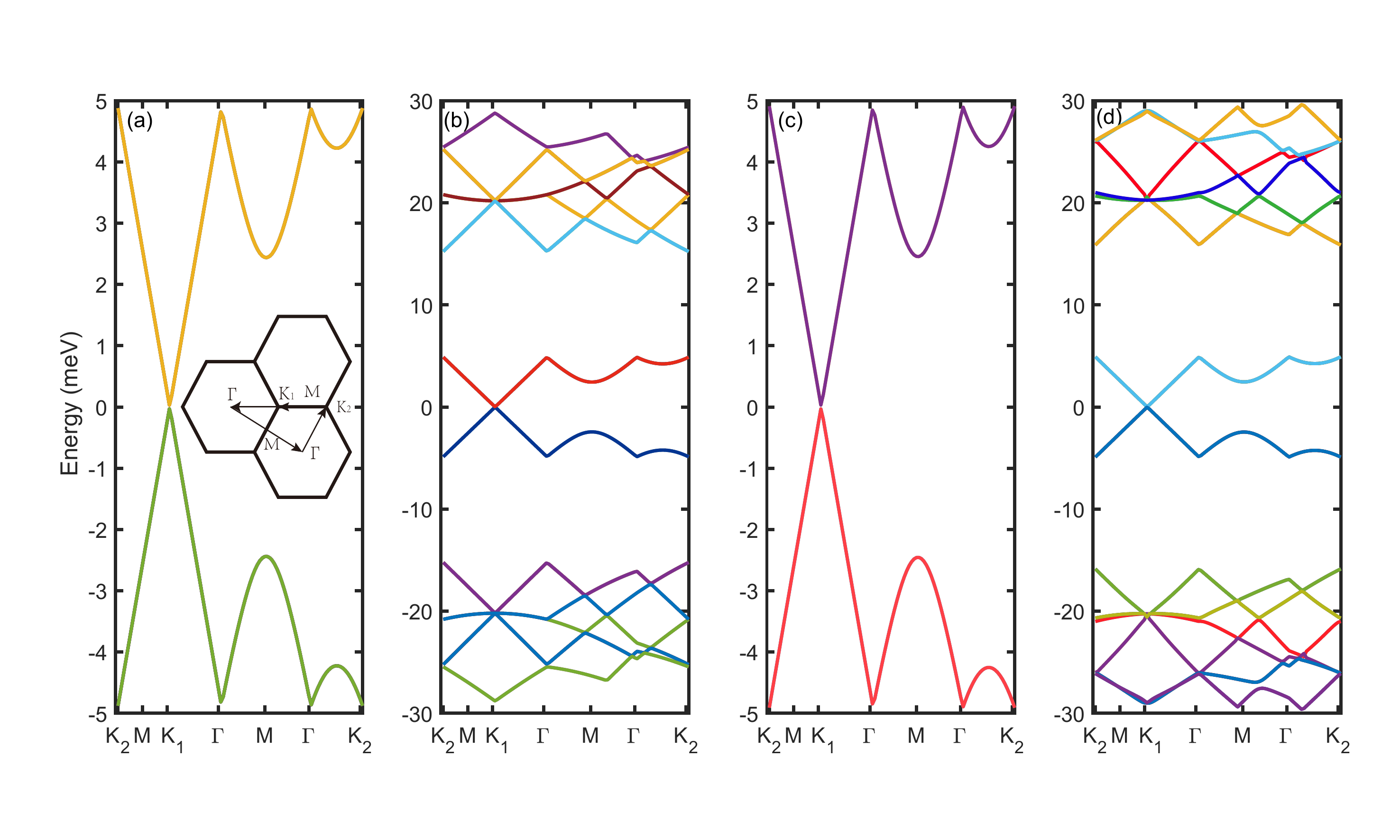}
		\caption{Flat band at magic angle $(\phi = 1.05^{\circ})$ based on Dirac equation [Eq.~\eqref{eq:hamiltonian}]. Bands closest to the Fermi surface ($\epsilon=0$) have been plotted. With $\Gamma=10^{-5}$ (a) Two-fold degenerate energy bands (four in total) and (b) Twenty-four bands. With $\Gamma=10^{-3}$ (c) Four bands and (d) 28 bands with size $L_{x}=15, L_{y}=15$. The mesh density is $N_{x}=N_{y}=15$. The band width is about $5~\text{meV}$. In subsequent figures, we show several bands around the Fermi energy unless otherwise specified. That is we assume the system is always at half filling. The origin of the {\textbf k} space is one of the moir\'e Dirac points, $\bf K_{1}$. Due to the non-Hermiticity and PT symmetry breaking nature of our model, the Dirac cone at the $\bf K_{2}$ point is gapped out.}
		\label{fig:band}
	\end{figure*}
	\begin{figure}[htp!]
		\includegraphics[width=1.05\columnwidth]{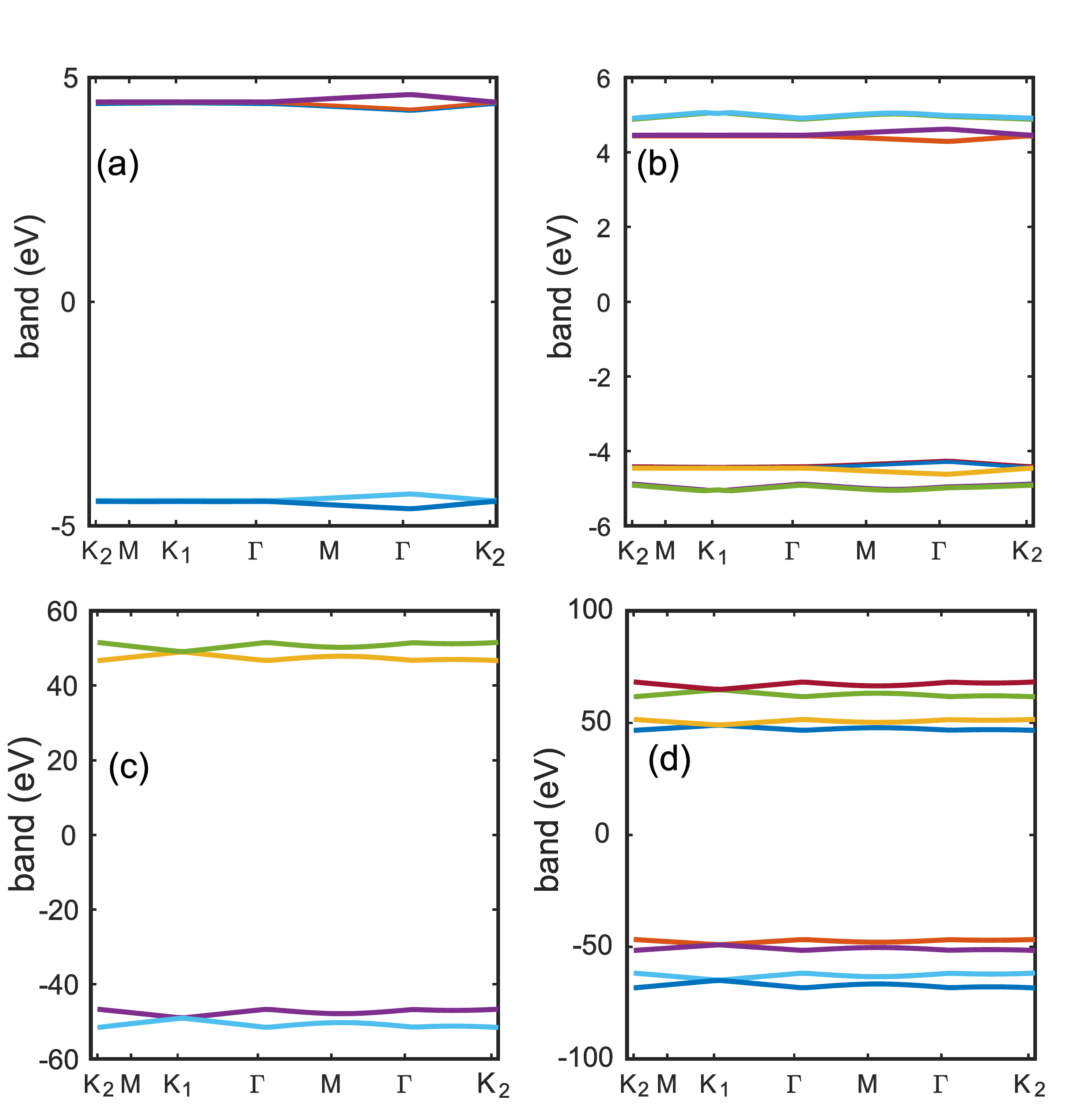}
		\caption{For Hermitian model, we show the result for 8 and 12 bands near $\epsilon=0$ with $\Gamma=10^{-3}$, $L_{x}=L_{y}=7$ in panels (a) and (b), respectively. In panels (c) and (d), we show the counterpart for parameter $\Gamma=10^{-5}$, $L_{x}=L_{y}=7$.  Mesh density $N_{x}=N_{y}=7$. Panel (c) shows 4 bands and Fig (d) shows 8 bands near $\epsilon=0$ respectively.}
		\label{fig:Hermitian}
	\end{figure}
	\begin{figure*}[htp!]
		\includegraphics[width=2\columnwidth]{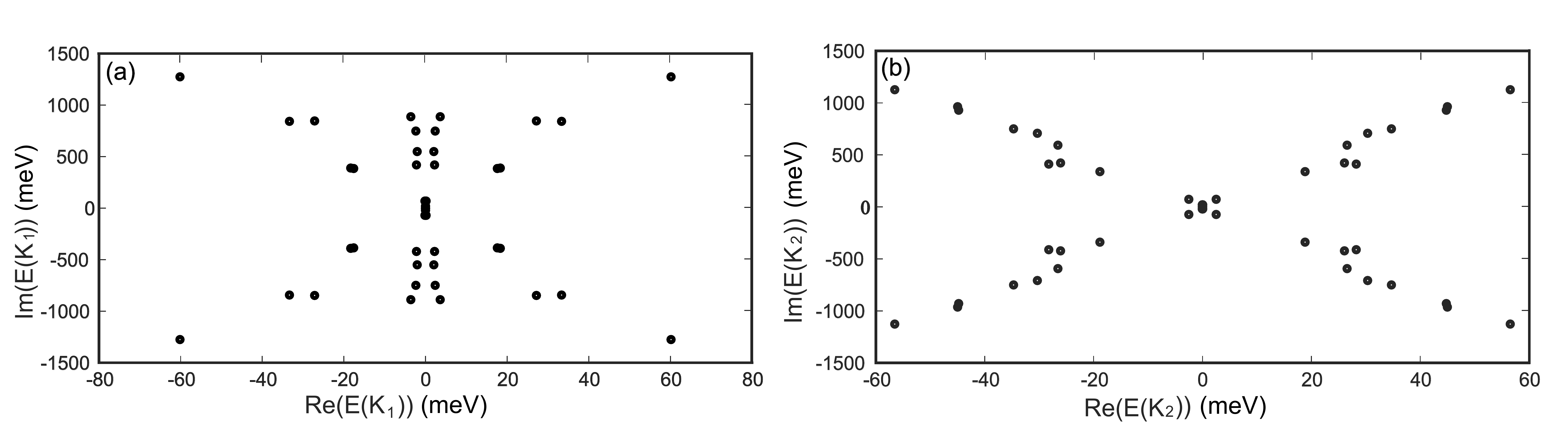}
		\caption{Real and imaginary values of the energy. (a) Illustrates the complex energy with $\Gamma=10^{-3}$, ~$L_{x}=L_{y}=7$ at $\textbf K_{1}$. (b) shows the complex energy with $\Gamma=10^{-3}$,~$L_{x}=L_{y}=7$ at $\textbf K_{2}$. Mesh density $N_{x}=N_{y}=7$. The spectra are symmetric about the real and imaginary axes.}\label{fig:complex}
	\end{figure*}
	
	\begin{figure*}[htp!]
		\includegraphics[width=2\columnwidth]{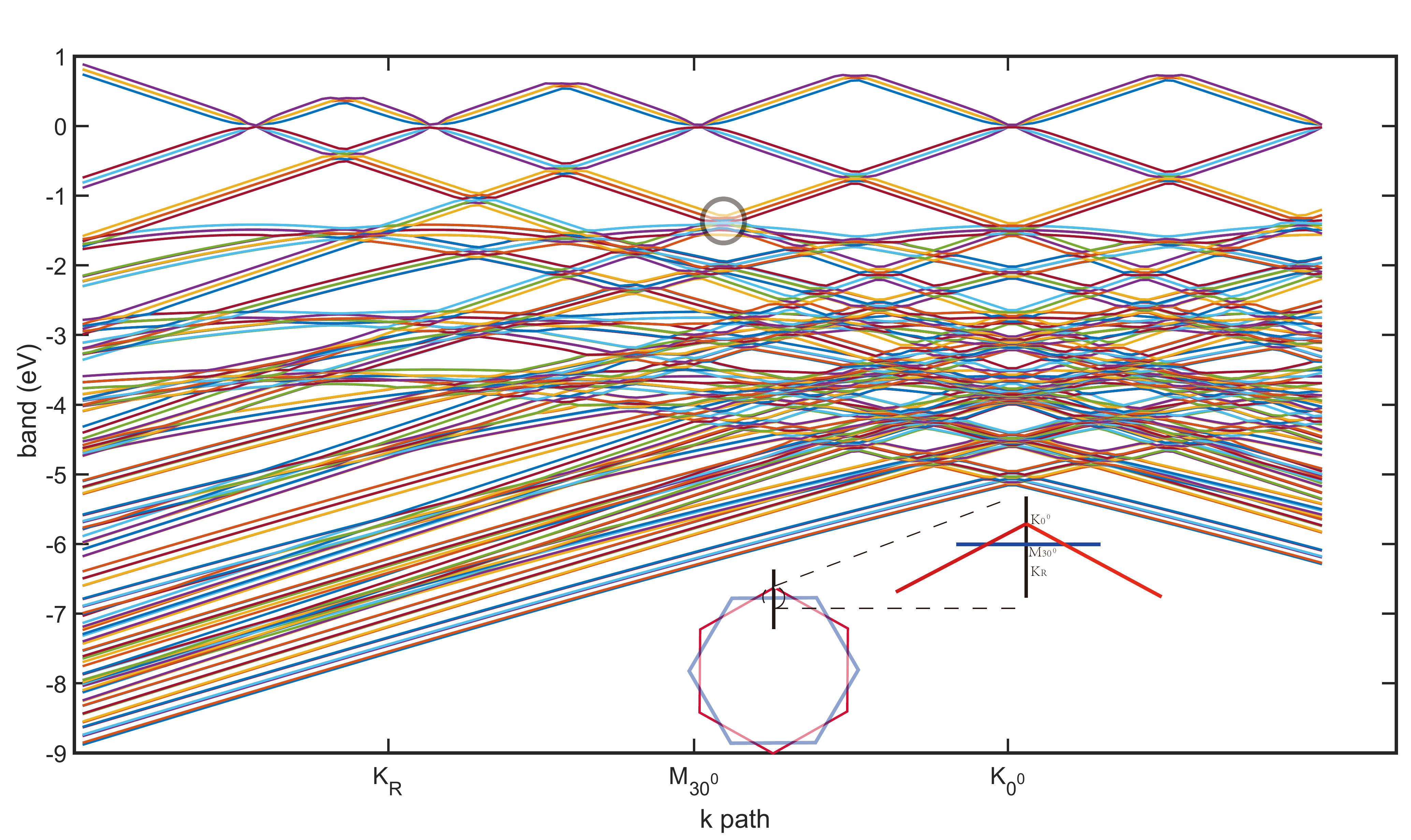}
		\caption{The 200 bands for $30^{0}$ QCTBG under $\epsilon=0$ with $\Gamma=0.5$. Other parameters are the same as Fig.~\ref{fig:band}. The band crossing in the black circle is the possible position for gap. In the minifigure, the red hexagon is the $0^{0}$ twisted (bottom) layer's BZ. The corresponding Dirac cone we denote as $K_{0^{0}}$. The blue counterpart is the $30^{0}$ twisted (top) layer. The M point of the $30^{0}$ layer's BZ is denoted by $M_{30^{0}}$. The mirror Dirac cone $K_{R}$ is symmetric with $K_{0^{0}}$ about the BZ boundary of $30^{0}$ layer. The momentum path is choosed to be the one connecting $K_{0^{0}}$ and $K_{R}$. The width of the bunch is close to ~$280~\text{meV}$.}
		\label{fig:QCTBGmirrorDirac}
	\end{figure*}
	In this section we elucidate how to get torsion free spin connection via vierbein and solve the relevant curved space Dirac equation with spin connection. First, we start with the flat metric under Cartesian coordinate.
	\begin{equation}
	ds^{2}=\eta_{ab}dx^{a}dx^{b}=-dt^{2}+dx^{2}+dy^{2}+dz^{2},
	\end{equation}
	where $\eta_{ab}=\text{diag}(-1,1,1,1)$. Assuming there is no torsion since we are considering rotation but no dislocation for the relative displacement between two layers the spin connection can be expressed using the vierbein. This will introduce a compact gauge field which characterizes the minimal coupling between effective curved space-time and pseudospin, that is the SU(2) gauge field or pseudo magnetic vector potential~\cite{Guinea2012prl,Jian-PengLiuXi-Dai2019prb,YaoWang2020prl}. This pseudo magnetic field can generate the strongly correlated flat bands and fractional quantum Hall effect (FQHE). The torsion free spin connection (expressed by flat background embedded space-time coordinate) is expressed as follows~\cite{Kleinert1989Gauge}
	\begin{equation}
	\begin{split}
	(\omega_{a})_{bc}=\frac{1}{2}\eta_{bd}\xi_{\mu}^{d}(\partial_{c}\partial_{a}-\partial_{a}\partial_{c})u^{\mu}+\frac{1}{2}\eta_{cd}\xi_{\mu}^{d}(\partial_{a}\partial_{b}-\partial_{b}\partial_{a})u^{\mu}\\+\frac{1}{2}\eta_{ad}\xi_{\mu}^{d}(\partial_{c}\partial_{b}-\partial_{b}\partial_{c})u^{\mu}+\frac{1}{2}(\partial_{a}\eta_{bc}+\partial_{c}\eta_{ba}-\partial_{b}\eta_{ca}). \label{eq:connection} \\
	\end{split}
	\end{equation}
	The vierbein and deformation field in Eq.~\eqref{eq:connection} can be obtained from Eq.~\eqref{eq:vierbein}. If the reference metric for an untwisted space-time is a flat one (like Euclidean or Minkowskian in our case), the spin connection can be reduced to the following form
	\begin{equation}
	\begin{split}
	(\omega_{a})_{bc}&=\frac{1}{2}\eta_{bd}\xi_{\mu}^{d}(\partial_{c}\partial_{a}-\partial_{a}\partial_{c})u^{\mu}+\frac{1}{2}\eta_{cd}\xi_{\mu}^{d}(\partial_{a}\partial_{b}-\partial_{b}\partial_{a})u^{\mu}\\&+\frac{1}{2}\eta_{ad}\xi_{\mu}^{d}(\partial_{c}\partial_{b}-\partial_{b}\partial_{c})u^{\mu}.\label{eq:connection1}
	\end{split}
	\end{equation}
	However when the background metric is curved, for example, cylindrical coordinate metric, the last term in Eq.~\eqref{eq:connection} will not vanish. One should also notice that only when the deformation field $u^{\mu}$ has a discontinuous second order mixed derivative, the deformation will produce a non-vanishing spin connection.
	\\
	\indent\setlength{\parindent}{1em}
	In principle, one can get the analytic form of the expression $(\partial_{x}\partial_{y}-\partial_{y}\partial_{x})\textbf{u}$ from   Eq.~\eqref{eq:deformationvectorA}~--~Eq.~\eqref{eq:deformationvector} for obtaining spin connection in Eq.~\eqref{eq:connection1}. However, obtaining the complete form of $(\partial_{x}\partial_{y}-\partial_{y}\partial_{x})\textbf{u}$ analytically is complicated, so in practice one should determine the expression numerically. The general term in $(\partial_{x}\partial_{y}-\partial_{y}\partial_{x})\textbf{u}$ is the product of the Dirac delta function, Heaviside function, floor function, and the derivative of the floor function. that is, $ (\partial_{x}\partial_{y}-\partial_{y}\partial_{x})\textbf{u}\approx \sum_{\langle 1,2,3\rangle}\textbf{A}_{123}\delta(\text{edge}_{1})\text{floor}'(\text{edge}_{1})\vartheta(\text{edge}_{2})\vartheta(\text{edge}_{3})$, where $\delta(x)$ is the Dirac delta function. To avoid numerical singularity we introduce a Lorentzian width $\Gamma$ for the domain wall in the spin connection [Eq.~\eqref{eq:connection1}] and the vierbein [Eq.~\eqref{eq:vierbein}], instead of using the Dirac delta function. $\Gamma$ is a small value with dimension of length, which implies broadening of the deformation due to some disorder, phonon or fluctuation (within a phenomenological analysis, this would be a tunable parameter) . More details can be found in Eq.~\eqref{eq:width} in Appendix ~\ref{sec:nonhermiticity}. 
	\\
	\indent\setlength{\parindent}{1em}
	We can obtain the total non-abelian connection by contracting its components with the pseudo spin-generators
	\begin{equation}
	\begin{split}
	\omega_{\mu}=\frac{i}{2}(\omega_{\mu})_{ab}\Sigma^{ab},\quad \Sigma^{ab}=\frac{i}{4}[\gamma^{a},\gamma^{b}],\quad
	\omega_{a}=e^{\mu}_{a}\omega_{\mu},\\
	\gamma^{0}=\left( \begin{array}{ccc} 0 & I \\ I & 0\end{array}\right),\quad
	\gamma^{i}=\left( \begin{array}{ccc} 0 & \sigma^{i} \\ -\sigma^{i} & 0\end{array}\right),\quad i=1,2,3,\label{eq:connection2}
	\end{split}
	\end{equation}
	where $[\gamma^{a},\gamma^{b}]=\gamma^{a}\gamma^{b}-\gamma^{b}\gamma^{a}$. The spin connection will modulate the hopping (for the tight binding model) and the covariant derivative (for the continuum model) as
	\begin{subequations}
		\begin{equation}
		\begin{split}
		t_{ij}&\to t_{ij}\exp\left(\frac{i}{4}dx_{ij}^{a}\left[\big{(}\omega_{a}(r_{i})\big{)}_{bc}+
		\big{(}\omega_{a}(r_{j})\big{)}_{bc}\right]\Sigma^{bc}\right),\\ \label{eq:hopping}
		\end{split}
		\end{equation}
		\begin{equation}
		\begin{split}
		\partial_{a}&\to D_{a}=\partial_{a}+\omega_{a}. \label{eq:derivative1}
		\end{split}
		\end{equation}
	\end{subequations}
	For the first expression in Eq.~\eqref{eq:derivative1}, the model corresponds to a non-abelian Hofstadter model with inhomogeneous flux~\cite{Hofstadter1976prb,cai2019arxiv,debeule2021prb,QuanShengWu2021prl}. We have omitted the pseudospin indices on the left-hand side in Eq.~\eqref{eq:hopping}. It is a diagonal matrix in the pseudospin and chirality direct product space. Since we are considering a continuous Dirac effective model, an alternate way to introduce the spin connection in Eq.~\eqref{eq:derivative1} will be implemented. \\
	\indent\setlength{\parindent}{1em}
	We consider a curved space Dirac action for TBG which has been studied in strained graphene system~\cite{Guinea2010PhysRep} and the Kitaev honeycomb lattice model with Kekul\'e distortion~\cite{Pachos2020prb}. Such a geometric theory has been applied to analyze gravitational chiral anomaly in the Weyl system~\cite{Nissinen2020prb,Nissinen2020prl}. Thus, we consider an effective theory at charge neutrality (half filling). Intuitively, this implies solving the Dirac fermion in the TBG curved space.
	
	The spin-$\frac{1}{2}$ quantum field in Riemannian space-time is governed by the following action and the Hamiltonian density (for TBG, for single valley approximation)~\cite{Pachos2020prb,golan2021arxiv,Xin-ChengXie2020prb}
	\begin{subequations}
		\begin{equation}
		\begin{split}
		S&=i\int d^{3+1}x|\xi|(\bar{\psi}\gamma^{\mu}D_{\mu}\psi+im\bar{\psi}\psi),\\
		\label{eq:action}
		\end{split}
		\end{equation}
		\begin{equation}
		\begin{split}
		H&=-i|\xi|[v_{f}(\overline{\psi}\gamma^{j}\partial_{j}\psi+\overline{\psi}\gamma^{j}\omega_{j}\psi)+im\overline{\psi}\psi],\\ \label{eq:hamiltonian}
		\end{split}
		\end{equation}
		\begin{equation}
		\begin{split}
		g_{\mu\nu}&=\frac{\partial x^{a}}{\partial y^{\mu}}\eta_{ab}\frac{\partial x^{b}}{\partial y^{\nu}}=
		\xi^{a}_{\mu}\eta_{ab}\xi^{b}_{\nu},
		\label{eq:metric}
		\end{split}
		\end{equation}
	\end{subequations}
	where\quad$|\xi|=|\det(\xi^{a}_{\mu})|=\sqrt{|\det(g_{\mu\nu})|}$, $D_{\mu}\psi=\partial_{\mu}\psi+\omega_{\mu}\psi$, $\mu=0,1,2,3,j=1,2,3$. The pseudospin spinor is given by $\psi=(\psi_{\uparrow R},\psi_{\downarrow R},\psi_{\uparrow L},\psi_{\downarrow L})^{T}$ where the $\uparrow,\downarrow$ means pseudospin. Chirality is indicated by $R$ and $L$. The $R$ spinor is a holomorphic function while the $L$ spinor is an anti-holomorphic function in a plane. The Dirac fermions are defined as usual by $\bar{\psi}\equiv \psi^{\dagger}\gamma^{0}, \gamma^{\mu}=e^{\mu}_{a}\gamma^{a}$. Unlike former theories which treat the layer as an internal degree of freedom and construct a 2+1d Dirac Hamiltonian, we use a 3+1d Dirac Hamiltonian. We consider the layer as a spatial coordinate. We assume ballistic scattering in the interlayer coupling direction so the bare Fermi velocity in $z$ is still $v_{f}$. If one considers the interlayer coupling as a tunneling or diffusive event, the Fermi velocity in the $z$ direction will be purely imaginary. However the nearly zero flat bands will still survive. For more details on this issue please refer to Appendix~ \ref{app:appc}.
	
	The interlayer coupling is controlled by two terms in Eq.~\eqref{eq:action}, $|\xi|(-iv_{f})\bar{\psi}\gamma^{z}\partial_{z}\psi$ and $-|\xi|\bar{\psi}w\psi$. The moir\'e modulation is hidden in the Jacobian $|\xi|$. In general, $|\xi|$ for AA/BB and AB/BA stacking region will be different. And the $|\xi|$ in AA/BB region will not necessarily be zero. So this model is beyond the chiral limit model~\cite{Ashvin2019prl}. One should also notice that although we don't consider both valleys, we start from a fully real space theory. Our vierbein theory can still capture the detailed short range information and the fast varied moire potential occurring in a large twisted angle setup.
	
	The spin connection in Eq.~\eqref{eq:hamiltonian} will be given by Eq.~\eqref{eq:connection2}. By introducing $\psi_{\textbf{k}}(\textbf{r})=\exp(i\textbf{k}\cdot\textbf{r})u_{\textbf{k}}(\textbf{r})$ and diagonalizing the Hamiltonian one can get the bands for arbitrary twisted angle. Even though $\textbf{k}$ is not a good quantum number in this case, one has to choose a $\textbf{k}$ in the background Euclidean space. The Dirac equation in the moir\'e scale (with about 1000 atoms per supercell) is computed using twisted boundary condition~\cite{YSWu1985prb,Bernevig2020science,HuangGuangYao2020prb}.
	\begin{equation}
	\psi_{\textbf{k}}(\textbf{r}+\textbf{L}_{r})=\exp\left(-\int_{\textbf{r}}^{\textbf{r}+
		\textbf{L}_{r}}ds~
	\omega_{r}\right)\psi_{\textbf{k}}(\textbf{r}). \quad r=x,y\label{eq:boundary}
	\end{equation}
	\indent\setlength{\parindent}{1em}
	We present a numerical solution~\cite{Boada2011IOP} for the Dirac equation in effective curved space (in a finite size $x\in[0,L_{x}], y\in[0,L_{y}]$, with the rotation axis located at (0,0). All length scales will be in units of $a_{0}=1.42$~\text{\r{A}}, the intralayer nearest distance between carbon atoms unless specified otherwise, at the first magic angle $\theta\approx 1.05^{0}$ as shown in Fig.~\ref{fig:band}. The discretization is implemented in the background Euclidean space with relevant spin connection and metric, which implies that we embed the curved manifold back to the background 3d Euclidean space~\cite{Fukaya2022arxiv}. We choose the parameters as follow. The bare Fermi velocity is $v_{f}=10^{6}m/s$. Interlayer distance is $h=0.5\times 1.42=0.71$~\text{\r{A}}. And the bare reduced interlayer interaction is $w=0.11~eV$ (acting as an effective mass) as in Ref.~\cite{MacDonald2011pnas}. The result of the lowest band width in Fig.~\ref{fig:band} is near the flat band width given by the BM model or DMRG (about 5 meV)~\cite{Senthil2019prb,Zaletel2020prb}. The nearly flat Dirac cone is located at K1 in moire BZ. Since this is a single valley theory, the lowest flat bands are gapped at other $\bf K$ points including $\bf K_{2}$. At the non-interacting level, MATBG is a narrow band semimetal. So this geometric theory can reproduce the flat bands given by former theory~\cite{Ashvin2019prl} qualitatively. Only the real part of bands has been shown in Fig.~\ref{fig:band}. 
	
	Our model introduces gaps in the band structure. First, even in the absence of relaxation, the lowest bands are gapped from the higher bands. Compared to the BM model and DFT calculations~\cite{HongMingWeng2022prb}, our real space model can capture the additional short range (UV) information than the $k$ space theories whose cutoff is $~1/a_{0}$, where $a_0$ is the nearest distance between carbon atoms in the monolayer. As shown in Fig.~\ref{fig:wholeplanedeformation}, our model can resolve the deformation within a length scale of $a_{0}$. This short range perturbation can gap out the lowest bands and higher ones. While in Ref.~\cite{HongMingWeng2022prb}, only when the relaxation is switched on, the lowest bands can be isolated from higher bands. We see that in Fig.~\ref{fig:wholeplanedeformation}, the deformation for a given point is pointing to the nearest untwisted site, which means if one introduces a Dirac fermion on a given lattice point it can feel the attraction from the nearest site. This is consistent with the tight-binding picture and the tendency for in-plane relaxation to occur. Second, due to the non-Hermiticity and PT symmetry breaking of our model, the Dirac cone at $\bf K_{2}$ in Fig.~\ref{fig:band} is also gapped out.\\
	\indent\setlength{\parindent}{1em}Additionally, we note that within a single valley BM model both $\bf K_{1}$ and $\bf K_{2}$ should be gapless. But the geometric theory presented here predicts that $\bf K_{2}$ is gapped. The reason may be as follows. In principle the dual momentum space should also be curved under our geometric theory. Thus, an appropriate moir\'e BZ should have been a curved one. However, we use a flat $k$- space to reduce the complexity of the problem. If one naively embeds the curved space into a flat one it may cause the spurious gaping of the high symmetry $\bf K_{2}$ point. The PT symmetry (analogous to the role of $C_{2}T$ symmetry in 2+1d TBG models) may also have been broken, causing the Dirac node to become gapped. Additionally, we have also computed the vierbein theory with multi Dirac nodes located at each mini-valley ($\bf K_{1}$ and $\bf K_{2}$) within a flat space-time theory in the absence of deformation. However, when we turn on the deformation the resulting lowest bands show gaps both at the $\bf K_{1}$ and $\bf K_{2}$ points with a gapless point appearing at $\bf M$. It will be an useful future exercise to study this feature of our model.
	
	\indent\setlength{\parindent}{1em}In a curved background, the Hamiltonian [Eq.~\eqref{eq:hamiltonian}] will generally be  non-Hermitian, as explained in detail in Appendix~\ref{sec:nonhermiticity}. Though the bands are usually complex, one can numerically verify that eigenvalues $a+ib,a-ib,-a+ib,-a-ib$ will simultaneously appear as shown in Fig.~\ref{fig:complex}. Very recently, Refs.~\cite{QiZhou2022natcomm,Kunst2021arxiv,ShaolongWan2021arxiv} have argued the duality relationship between non-Hermitian model in flat space and Hermitian system in curved space. So treating a non-Hermitian Dirac action is justified. In fact, one can always construct a Hermitian Hamiltonian by substituting $S\to \frac{1}{2}(S+S^{\dagger}), H\to\frac{1}{2}(H+H^{\dagger})$ as in Ref.~\cite{Nguyen2020arxiv}. However, in order to ensure the existence of nearly zero energy bands, one should introduce the covariant derivation operator $-iD_{\mu}$ in the action and not $\frac{1}{2}(-iD_{\mu}+i\overline{D}_{\mu})$. The existence of zero energy band in the deformed bilayer graphene is ensured by the
	Atiyah-Singer index theorem~\cite{Prokhorova2008prb}. Thus, we focus on the non-Hermitian Dirac Hamiltonian.
	
	In Fig.~\ref{fig:Hermitian}, we observe that there are no energy bands around the Fermi surface for the Hermitian Hamiltonian. The lowest bands are no longer as flat as its non-Hermitian counterpart. Since the magnitude of bandwidth in Fig.~\ref{fig:Hermitian}(a) and \ref{fig:Hermitian}(b) is at least 0.1 eV, if one decreases the domain wall width to $\Gamma=10^{-5}$ (in the unit of $a_{0}$) as in Fig.~\ref{fig:Hermitian}(c) and \ref{fig:Hermitian}(d), the energy scale of the lowest bands will increase by an order of magnitude. Whereas in Fig.~\ref{fig:band}, the domain wall width will not significantly influence the lowest band structure. The crucial flat bands still survive. Besides, there is an alternative way to argue that the TBG Hamiltonian should be non-Hermitian.
	
	One can design an adiabatic process for twisting the bilayer from AA stacking to one of the magic angles. If the  TBG Hamiltonian is Hermitian for all twisted angle, the wave function for AA stacking and magic angle can be connected by a local unitary transformation (adiabatic time evolution). However, the AA stacking bilayer graphene is topologically trivial while MATBG hosts a stable topological phase~\cite{Bernevig2019prl}. So during the adiabatic twisting process, there must be a topological phase transition. Thus, the assumption for adiabatic unitary evolution is incorrect~\cite{XGWen2010prb,Budich2021prr}. (For realistic situation with a hBN substrate, one can expect that there will always be a gap between the ground state and the excited state for initial and final wave function.) There should be at least one instant of a non-Hermitian Hamiltonian during the adiabatic twisting process. One can observe that for the lowest energy levels in Fig.~\ref{fig:complex}, the imaginary part has the same magnitude as the real part. So the non-Hermiticity will not cause apparent broadening for the lowest band's density of state. Such a spectrum is protected by the so-called bi-chiral symmetry in certain parameter regime~\cite{cai2019arxiv}. The discontinuity of $\frac{\partial E_{\textbf{k}}}{\partial k}$ at $\bf\Gamma,K_{2}$ in Fig.~\ref{fig:band} can be interpreted as the anisotropy in the Fermi velocity at $ \bf\Gamma,K_{2}$~\cite{Shaffique2019pnas} in the moir\'e BZ since at these {\bf k} points the {\bf k} path changes direction.
	\\
	\indent\setlength{\parindent}{1em}We end this section by demonstrating how our theory is applicable to an incommensurate system. We compute the bands for a $30^{0}$ quasicrystal TBG (QCTBG)~\cite{HaiLinPeng2020acsnano,ShuYunZhou2018pnas,YoungWoo2019prb}. We reproduce the mirror Dirac cone in Ref.~\cite{ShuYunZhou2018pnas}. The occurrence of a mirror Dirac cone $K_{R}$ in the Raman signal requires the presence of phonons. Thus, we choose a relatively large domain wall width $\Gamma=0.5$ during the simulation. In our coordinate system, the red hexagon in the minifigure in Fig.~\ref{fig:QCTBGmirrorDirac} is the $0^{\circ}$ twisted (bottom) layer's Brillouin zone (BZ). The corresponding Dirac cone we denote as $\bf K_{0^{\circ}}$. The blue counterpart is the $30^{\circ}$ twisted (top) layer. The M point of the $30^{\circ}$ layer's BZ is denoted by $\bf M_{30^{\circ}}$. The mirror Dirac cone $\bf K_{R}$ is symmetric with $\bf K_{0^{\circ}}$ about the BZ boundary of the $30^{\circ}$ layer. The momentum path is chosen to be the one connecting $\bf K_{0^{\circ}}$ and $\bf K_{R}$. The Raman and ARPES experiments in Fig.~3 from Ref.~\cite{ShuYunZhou2018pnas} verifies that there is a gap at $\bf M_{30^{\circ}}$. Thus, the result based on our theory qualitatively reproduces the mirror Dirac cone. The black circle in Fig.~\ref{fig:QCTBGmirrorDirac} shows the position of the possible gap. There are a bunch of bands crossing the black circle. The width of the bunch is close to the gap size $280meV$ as in Ref.~\cite{ShuYunZhou2018pnas} (Fig.5). However note that our theory is a non-interacting single valley theory. So one may introduce interaction, interband hybridization and inter-valley coupling to get the correct gap size at $\bf M_{30^{\circ}}$, which is worthy of future study.
	\\
	
	\section{Theory of rotating bilayer graphene (RBG)}
	\label{sec:RBG}	
	\begin{figure*}[htp!]
		\includegraphics[width=2\columnwidth]{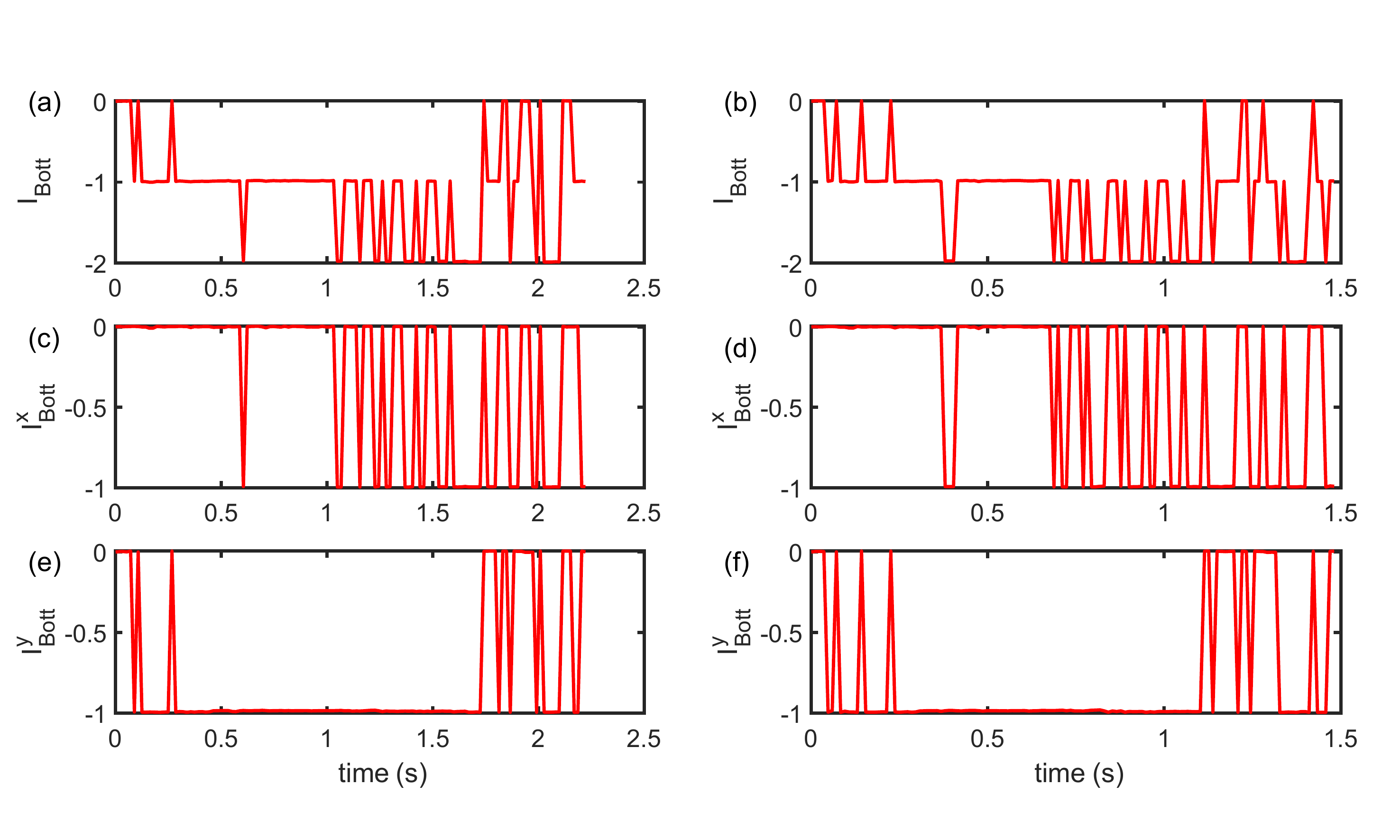}
		\caption{Bott index in one period when $\Gamma=0.5$ with size $L_{x}=15, L_{y}=15\sqrt{3}$. The thickness of monolayer graphene is $h=2.36$. The mesh density $N_{x}=N_{y}=15$. Panels (a), (c), and (e) are the total Bott index, Bott index in the $x$ direction, and the $y$ direction  [Eq.~\eqref{eq:Bott}] under 0.943 rad/s of driving frequency, respectively. While (b), (d), and (f) are for 1.414 rad/s of driven frequency. For each diagram, the number of the time slices is 125, that is, the time step is $T/125$.}
		\label{fig:Bott}
	\end{figure*}
	\begin{figure*}[htp!]
		\includegraphics[width=2\columnwidth]{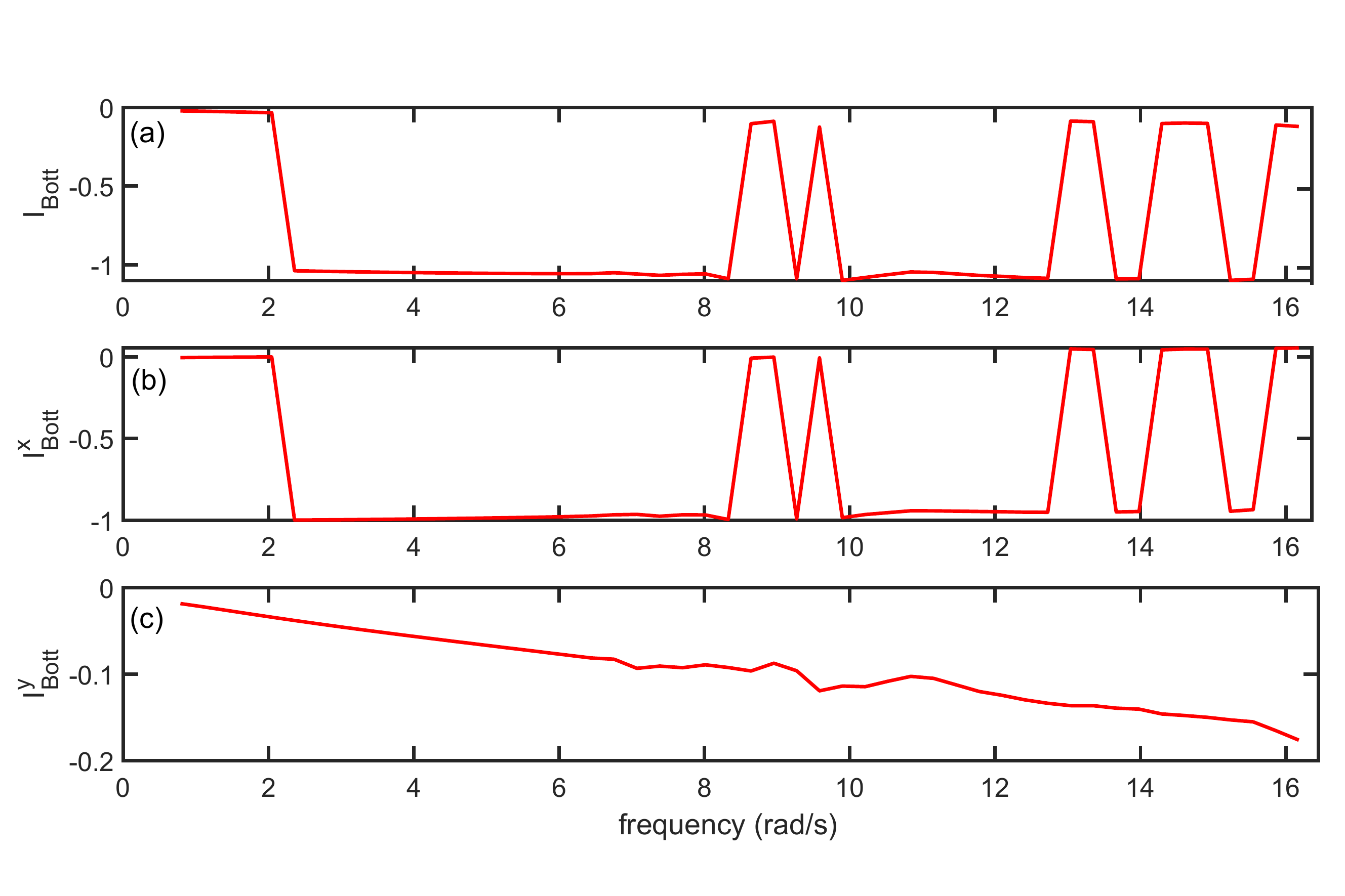}
		\caption{Bott index-driving frequency relation $I(\omega,T)$ when $\Gamma=0.5$ with size $L_{x}=15, L_{y}=15\sqrt{3}, h=2.36$. The mesh density is $N_{x}=N_{y}=15$. Fig.~(a),~(b),~(c) are $I_{Bott},~I^{x}_{Bott},~I^{y}_{Bott}$ respectively. The frequency step in simulation is 0.314rad/s. The Bott indices for 50 different driven frequency are shown.}
		\label{fig:Bottf}
	\end{figure*}
	\begin{figure}[htp!]
		\includegraphics[width=1\columnwidth]{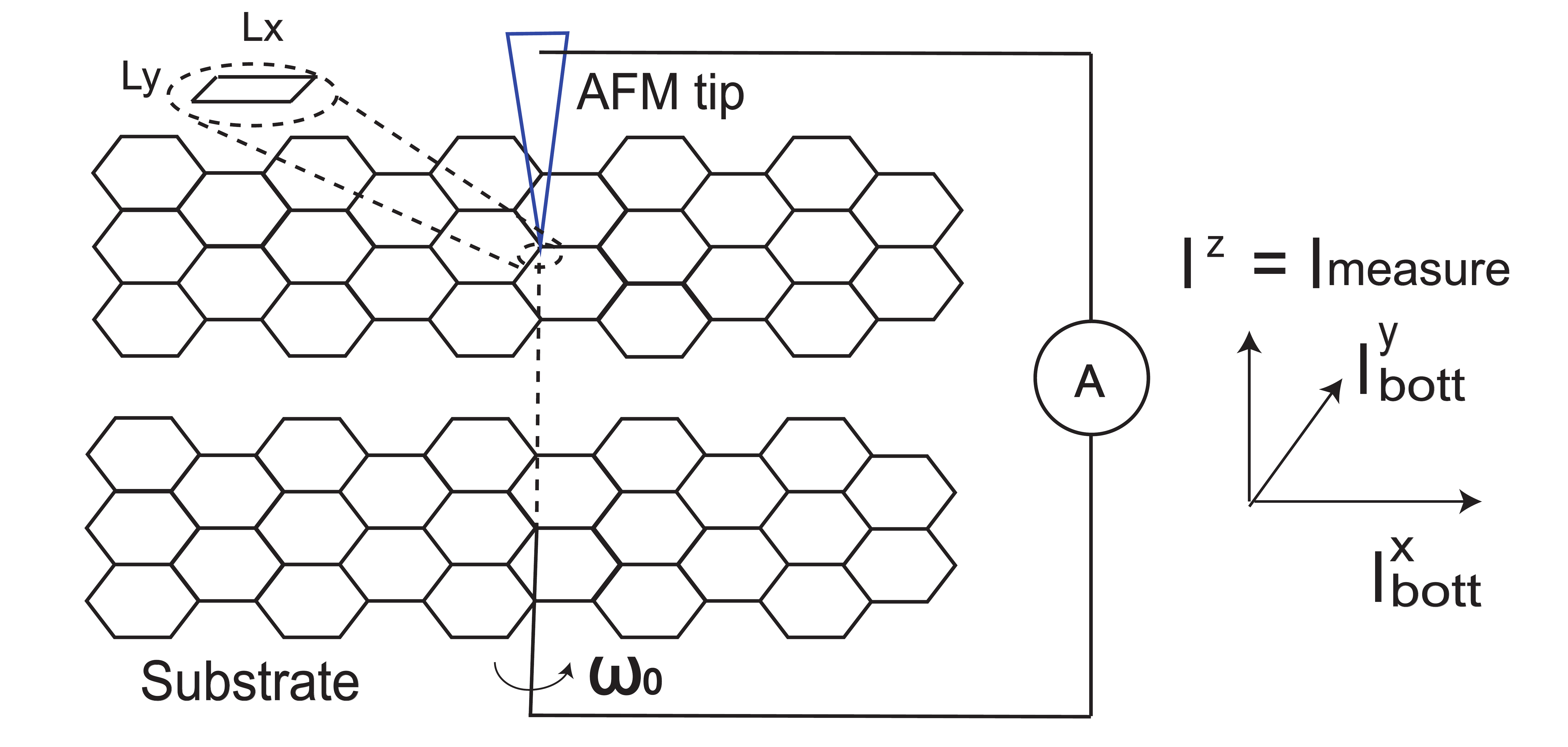}
		\caption{A simple sketch for realizing the measurement of RBG pumping current. In practice, rotating the subsrtate is more convenient. The total transverse pumping current can be measured by the vertical ammeter when the system achieves a stable state. The Atomic Force Microscope (AFM) tip is tailored according to the given finite size $L_{x}\times L_{y}$ rectangle.}
		\label{fig:experiment}
	\end{figure}
	\setlength{\parindent}{2em}In this section we generalize our geometric theory from TBG to RBG. This is the relative rotating bilayer graphene configuration with constant angular velocity $\omega_{0}$. The motivation to investigate RBG is that under the vierbein formalism, a RBG will generate a spin connection with non-vanishing time component $\omega_{t}$, which can mix with the usual Aharonov-Anandan phase (generated by periodic evolution)~\cite{Aharonov1987prb}. As a result new topological phase or phases will arise as evidenced from the non-trivial Bott index regime in Fig.~\ref{fig:Bottf}. Another motivation is that although Floquet engineering of TBG has been systematically studied~\cite{Fiete2020prb,Torma2021prb,LedeXian2019prr,Shun-YuYao2017prb}, most of them have focused on light frequency driven regime or sliding TBG as a 2d Thouless pump~\cite{Thouless1983prb,DiXiao2020prb}. The non-adiabatic structurally rotating driving TBG is still in need of investigation~\cite{Chew2020prb,Clemens2022arxiv}. The non-adiabatic Thouless pump has only been reported in waveguide array ~\cite{Fedorova2020natcomm}. Actually, the vierbein formalism can be naturally generalized to RBG. The only difference is that one should use a four dimensional metric and let $\phi=\omega_{0}t$. The Hamiltonian for time-dependent curved space-time Dirac fermion is written as~\cite{Boada2011IOP}
	\begin{equation}
	\begin{split}
	H=-i|\xi|[-\overline{\psi}\gamma^{0}\omega_{t}\psi+v_{f}(\overline{\psi}\gamma^{j}\partial_{j}\psi+\overline{\psi}\gamma^{j}\omega_{j}\psi)+im\overline{\psi}\psi].
	\label{eq:Ht}
	\end{split}
	\end{equation}
	\indent\setlength{\parindent}{1em}When the bilayer graphene starts rotating, another factor which will cause non-Hermiticity is the boost generator $\Sigma^{0j}=\frac{i}{4}[\gamma^{0},\gamma^{j}], (j=1\cdots 3)$ in Eq.~\eqref{eq:connection2}. It is an anti-Hermitian operator under dynamic vierbein situation in RBG. As a result, the gauge transformation factor in (dynamic) twisted boundary condition [Eq.~\eqref{eq:boundary}] will be a non-unitary one, which can be traced back to the non-compact nature of the Poincar\'e group~\cite{Ryder1985QFT}. So the system turns out to be a non-Hermitian Floquet system. One should also be cautious that for a general twist angle there is no moir\'e BZ. Since for a general incommensurate twisted angle, there is no strict translation symmetry our vierbein formalism in the {\bf k} space will also be a curved one for general incommensurate twisted angle~\cite{Bernevig2019prl,mera2021prb}. So in the following calculations, we will manipulate in real space instead of {\bf k} space.
	\\
	\indent\setlength{\parindent}{1em}
	To capture the quasiperiodic or incommensurate nature of RBG, we consider the Bott index and relate it to the charge pumping during rotation driving~\cite{Yoshii2021prb,YongXu2020prb,FengLiu2018prl,DongHuiXu2021prb,Ueda2018prx,ZQZhang2021prb,Zhi-QiangZhang2020cpb}.
	\begin{subequations}
		\begin{equation}
		\begin{split}
		P(t)&=\sum_{\text{Re}(E_{n}(t))<\mu}|\psi_{n}(t)\rangle \langle\tilde{\psi_{n}}(t)|,\\
		U_{r}&=I-P+P\exp\left(i2\pi\frac{r}{L_{r}}\right)P,\quad r=x,y.
		\label{eq:project}
		\end{split}
		\end{equation}
		
		\begin{equation}
		\begin{split}
		I_{Bott}^{r}(t)&=\frac{1}{2\pi}\int_{0}^{t}\text{Im}[\text{Tr}\Big{(}\ln\big{(}U_{r}(t'+d t')U_{-r}(t')\big{)}\Big{)}]\\
		&=\frac{1}{2\pi}\int_{0}^{t} dt'\partial_{t'}[\text{Im}\Bigg{(}\ln\Big{(}\det\big{(}U_{r}(t')\big{)}\Big{)}\Bigg{)}].  \quad r=x,y.
		\end{split}
		\end{equation}
		
		\begin{equation}
		\begin{split}
		I_{Bott}&=I_{Bott}^{x}+I_{Bott}^{y}. \label{eq:Bott}
		\end{split}
		\end{equation}
	\end{subequations}
	Let $V$ be the eigenstate matrix of the Hamiltonian. Each column corresponds to the eigenstate $\{|\psi_{n}\rangle\}$, while $\{|\tilde{\psi}_{n}\rangle\}$ in Eq.~\eqref{eq:project} is the counterpart for matrix $(V^{-1})^{\dagger}$ (biorthogonal basis~\cite{Brody2013IOP,HongWu2020prb}). $I_{Bott}(T)$ can be considered as the charge pumping per period for a given finite size $L_{x}\times L_{y}$ rectangle region~\cite{Yoshii2021prb,Thouless1983prb}. The factor $U_{x}(t'+d t')U_{-x}(t')$ contains the effect of the Aharonov-Anandan phase since it relates the gauge connection defining on the bundle whose base manifold isa  1d periodic time parameter in $~S^{1}$. 
	\\
	\indent\setlength{\parindent}{1em}
	The result of the RBG Bott index is shown in Fig.~\ref{fig:Bott} which shows the total contribution along with the individual contributions in the $x$ and the $y$ direction. The figure suggests that the Bott index can transition between trivial to non-trivial topological sectors. We observe from the figures that the Bott index variation is similar over the time cycle displayed. This variation guides us to the instantaneous time dependence of the Bott index. Thus, to obtain information on any potential topological transition, we have to study the behavior of this index across various frequency cycles.
	\\
	\indent\setlength{\parindent}{1em}
	In Fig.~\ref{fig:Bottf}, we show the dependence of Bott index $I(\omega_{0},T)$ on driven frequency $\omega_{0}$. We can see that in certain driven frequency regime, the finite size $L_{x}=15, L_{y}=15\sqrt{3}$ RBG system will hold non-trivial Bott index. Note that here we still use Cartesian coordinate instead of the hexagonal coordinate used in~\cite{Geim2009rmp}. So the structural rotation periodic driving may cause new topological phase. During the simulation, one assumes that the RBG system stays in half filling all the time. As bilayer graphene starts rotating and achieves steady state it will act as a non-adiabatic Thouless pump~\cite{Fedorova2020natcomm}. In one period, the charge pumping will be quantized and can be recorded by an ammeter. Numerical result shows that the quantized charge pumping will generally depend on the size and the driving frequency. To show the relation between Bott index and pumping charge consider the expectation of a wave packet center which can be expressed as~\cite{Yoshii2021prb},
	\begin{equation}
	\langle r(t)\rangle=\langle\psi(t)|\hat{r}|\psi(t)\rangle=\frac{L_{r}}{2\pi}\text{Arg}\left(U_{r}(t)\right),\quad r=x,y.
	\end{equation}
	In the above we used the abbreviation $\text{Arg}=\text{Im}\text{Tr}\text{ln}$. The corresponding polarized current pumping in an infinitesimal time interval can be related to the Bott index as
	\begin{equation}
	\begin{split}
	j_{r}(t)\Delta t=\text{Arg}\left(U_{r}(t+\Delta t)U_{-r}(t)\right),\\
	I_{Bott}^{r}(T)=\int_{0}^{T}dt j_{r}(t).\quad r=x,y.
	\end{split}
	\end{equation}
	To some extent, the Bott index can be interpreted as a finite size version of Hall conductivity or Chern number. As an analogy, one can consider the voltage caused by rotation as a kind of Faraday voltage from the change of the pseudo magnetic flux, $U=\frac{d\Phi}{dt}\approx \omega_{0}A$. And the differential Hall conductivity is defined as $\sigma_{H}=C\frac{e^{2}}{h}=\frac{dI}{dU}\approx \frac{dI^{Bott}}{d\omega_{0}}$. The step transition of the Bott index can be interpreted as the change of the finite size Chern number or Hall conductivity. This argument is similar to the Streda formula\cite{Streda1982IOP}.
	Thus the Bott index is directly connected to real space polarized pump current which can be measured directly. When the RBG achieves a stable state we have the following relationship amongst the current. $\partial_{t}Q+I_{x}+I_{y}+I_{z}=0, ~\partial_{t}Q=0,~I_{x}+I_{y}=I_{Bott}=-I_{z}=-I_{measure}$. So the total pumping current (Bott index) can be measured by a vertical current.
	\section{Discussion and conclusion}
	\label{sec:conclude}
	\indent\setlength{\parindent}{1em} We have developed a geometric effective theory for TBG with a generic twist angle. In principle this theoretical formulation can be generalized to other twisted bilayer Bravais lattice as long as the dual lattice is known and the atoms are homogeneous. For a TBG, by connecting a given location (the point where we want to obtain the deformation field) and the nearest twisted triangle center we can get the deformation field for an arbitrary twist angle and arbitrary position. This deformation can induce an effective SU(2) gauge field and an emergent curved space. The novel properties of the TBG can be interpreted as a geometric response. The numerical result of energy bands near the first magic angle for the non-Hermitian Dirac Hamiltonian reproduces the flat bands. The discontinuity of $\partial{E_{\textbf{k}}}/\partial k$ indicates the Fermi velocity anisotropy at high symmetry {\bf k}-points $\bf\Gamma$ and $\bf K_{2}$. When one uses imaginary vertical Fermi velocity model to simulate the TBG system, it emphasizes the decay in the wave function in the vertical direction. Numerically, the lowest flat bands still survive. We also show that our theory is applicable to the $30^{\circ}$ QCTBG (within limitations).
	
	We also showed that our effective geometric theory is applicable to the system with a dynamic vierbein such as a RBG. For a RBG with constant angular velocity $\omega_{\circ}$, the quantized pumping charge is illustrated by the Bott index which may be examined by transport experiment. A simple experimental proposal is also discussed to test the validity of the proposed Bott index theory for RBG. The topological property of RBG is controlled by the mixture of spin connection and the Aharonov-Anandan phase. Furthermore, note that the effective theory proposed in this article is still a single body theory, which does not take into account many body interaction. Thus, an explanation of the correlated insulating phase and the superconducting phase is beyond the present scope of the TBG and the RBG formalism. Finally we suggest an experimental setup shown in Fig.~\ref{fig:experiment} which can test the validity of our proposed theoretical formulation. In this setup we consider the bilayer system being probed by an Atomic Force Microscopy (AFM) setup where the tip which is tailored according to the given finite size of the system. With one of the layers fixed (say the upper one), the bottom is rotated, this should generate the transverse currents which can be measured within a transport setup. 
	
	\section*{Acknowledgements}
	We thank Zhong-Bo Yan, Peng Ye, Jian-Peng Liu, Yi-Wen Pan, Jie Ren, Shi-Dong Liang, Guo-Yi Zhu, Jiannis K. Pachos, Matthew Horner, Jia-Qi Cai, Zi-Ang Hu, Wen-Jie Xi, Ze-Min Huang, Jaakko Nissinen, Hong Wu, Guang-Jie Li, Yu-Han Liu, Zhi-Qiang Zhang, Zi-Jian Xiong, Jian-Keng Yuan, Ge-Wei Chen, Ding-Kun Lian, Jun Li, Yun-Feng Chen for valuable discussions. J. Z. M. and D. X. Y. are supported by NKRDPC-2017YFA0206203, NKRDPC-2018YFA0306001, NSFC-92165204, NSFC-11974432, GBABRF-2019A1515011337, Leading Talent Program of Guangdong Special Projects (201626003), and Shenzhen International Quantum Academy (Grant No. SIQA202102). T. D. acknowledges funding support from Sun Yat-Sen University Grants No. OEMT-2019-KF-04 and No. OEMT-2017-KF-06. TD acknowledges the hospitality of KITP at UC-Santa Barbara. A part of this research was completed at KITP and was supported in part by the National Science Foundation under Grant No. NSF PHY-1748958. 
	\appendix
	\section{Numerical implementation and non-Hermiticity}
	\label{sec:nonhermiticity}
	The discretization of the ordinary partial (momentum) operator in curved space-time is given by
	\begin{equation}
	\begin{split}
	&-i|\xi|\partial_{x}(j)\to -i|\xi_{j}|\left[\frac{\delta_{j,j+1}-\delta_{j,j-1}}{2\Delta x}\right],\\
	&-i|\xi|\partial_{x}(j+1)\to -i|\xi_{j+1}|\left[\frac{\delta_{j+1,j+2}-\delta_{j+1,j}}{2\Delta x}\right],
	\end{split}
	\end{equation}
	where $\delta$ is the Kronecker symbol and $j$ labels mesh site. $\Delta x$ means the differential step. Apparently, if the vierbein has space-time dependence, i.e $|\xi_{j}|\neq |\xi_{j+1}|$ generally, $ -i|\xi|\partial_{x}$ will be a non-Hermitian operator, i.e, $-i|\xi_{j}|\frac{\delta_{j,j+1}}{2\Delta x}\neq \left(i|\xi_{j+1}|\frac{\delta_{j+1,j}}{2\Delta x}\right)^{\dagger}$. If one imposes the hermiticity, one should substitute $ -i|\xi|\partial_{x}\psi\to -i|\xi|\left(\frac{1}{\sqrt{|\xi|}}\partial_{x}[\sqrt{|\xi|}\psi]\right)$ \cite{Barros2005EPJC} or choose the Hermitian Hamiltonian. The discretization according to the former strategy takes the form
	\begin{equation}
	\begin{split}
	&-i|\xi|\left(\frac{1}{\sqrt{|\xi|}}\partial_{x}[\sqrt{|\xi|}(j)]\right)\to \\ &-i\left[\frac{\sqrt{|\xi_{j+1}||\xi_{j}|}\delta_{j,j+1}-\sqrt{|\xi_{j}||\xi_{j-1}|}\delta_{j,j-1}}{2\Delta x}\right],\\
	&-i|\xi|\left(\frac{1}{\sqrt{|\xi|}}\partial_{x}[\sqrt{|\xi|}(j+1)]\right)\to\\ &-i\left[\frac{\sqrt{|\xi_{j+2}||\xi_{j+1}|}\delta_{j+1,j+2}-\sqrt{|\xi_{j+1}||\xi_{j}|}\delta_{j+1,j}}{2\Delta x}\right].
	\end{split}
	\end{equation} \\
	Here the non-Hermitian discretization has been implemented to reproduce flat bands.
	Physically, momentum operators in curved space or incommensurate system are generally
	non-Hermitian. As an example, we consider one dimension $-i\int^{+\infty}_{-\infty}dx\psi^{\dagger}\partial_{x}\psi=-i(\psi^{\dagger}\psi|^{+\infty}_{-\infty}-\int^{+\infty}_{-\infty}dx(\partial_{x}\psi^{\dagger})\psi)$. When $\psi^{\dagger}\psi|^{+\infty}_{-\infty}\neq 0$, momentum operators will be non-Hermitian. To do the numerical calculation, one should substitute the following functions in non-singular form when evaluating the vierbein [Eq.~\eqref{eq:vierbein}], spin connection [Eq.~\eqref{eq:connection1}], and metric [Eq.~\eqref{eq:metric}].
	
	\begin{equation}
	\begin{aligned}
	\delta(x)&\approx\frac{\Gamma}{\pi}\frac{1}{x^{2}+\Gamma^{2}},~~~~~~
	\vartheta(x)\approx\frac{1}{\pi}\arctan\left(\frac{x}{\Gamma}\right)+\frac{1}{2},\\
	\text{floor}'(x)&=\sum_{k=-\infty}^{\infty}\delta(x-k)\approx\sum_{k=-N}^{N}
	\frac{\Gamma}{\pi}\frac{1}{(x-k)^{2}+\Gamma^{2}}, \label{eq:width}
	\end{aligned}
	\end{equation}
	where $\Gamma$ is a small value with the dimension of length, representing the broadening of the deformation domain wall due to some disorder, phonon or fluctuation. Even when
	the system is very clean there is still an intrinsic quantum fluctuation (like phonon zero point energy). So, in general the domain wall broadening is inevitable. Here $N$ is an integer which is large enough and determined by the twist angle and size.
	\section{Second order derivative of rotating bilayer graphene}
	\label{sec:detail}
	For RBG, one should consider more components of spin connection. However one should be cautious that the partial operator for twisted angle is not independent of the spatial partial operator. One has the following expressions
	\begin{equation}
	\begin{split}
	&\phi=\omega_{0}t,\partial_{t}=\omega_{0}\partial_{\phi},\\
	&\partial_{\phi}=x\partial_{y}-y\partial_{x}=x_{2}\partial_{x_{1}}-x_{1}\partial_{x_{2}},\\
	&x_{1}=x\cos\phi+y\sin\phi,~x_{2}=-x\sin\phi+y\cos\phi.
	\end{split}
	\end{equation}
	Our target is to get the dynamic spin connection [Eq.~\eqref{eq:connection1}] in Eq.~\eqref{eq:Ht}. In the basis of ${x_{1},x_{2}}$, we obtain
	\begin{subequations}
		\begin{equation}
		\begin{split}
		(\partial_{\phi}\partial_{x}-\partial_{x}\partial_{\phi})
		\left(\begin{array}{cccc}u^{x}\\u^{y}\end{array}\right)=
		\\ \left[-A_{\phi}C_{\phi}(\partial_{x_{1}}\partial_{x_{2}}-\partial_{x_{2}}\partial_{x_{1}})
		-\frac{1}{2}E_{\phi}\partial_{x_{1}}-\frac{1}{2}D_{\phi}\partial_{x_{2}}\right]\left(\begin{array}{cccc}u'^{x}\\u'^{y}\end{array}\right),\\
		\end{split}
		\end{equation}
		\\
		\begin{equation}
		\begin{split}
		(\partial_{\phi}\partial_{y}-\partial_{y}\partial_{\phi})\left(\begin{array}{cccc}u^{x}\\u^{y}\end{array}\right)=\\ \left[-B_{\phi}C_{\phi}(\partial_{x_{1}}\partial_{x_{2}}-\partial_{x_{2}}\partial_{x_{1}})
		+\frac{1}{2}D_{\phi}\partial_{x_{1}}-\frac{1}{2}E_{\phi}\partial_{x_{2}}\right]\left(\begin{array}{cccc}u'^{x}\\u'^{y}\end{array}\right),\\
		\end{split}
		\end{equation}
	\end{subequations}
	
	\begin{equation}
	\begin{split}
	&A_{\phi}=x_{1}\cos\phi -x_{2}\sin\phi, \quad
	B_{\phi}=x_{1}\sin\phi+x_{2}\cos\phi,\\
	&C_{\phi}=\left( \begin{array}{cccc}\cos\phi&-\sin\phi\\ \sin\phi&\cos\phi\end{array}\right),\quad
	D_{\phi}=\left( \begin{array}{cccc}1+\cos 2\phi&-\sin 2\phi\\
	\sin 2\phi&1+\cos 2\phi\end{array}\right), \\
	&E_{\phi}=\left(\begin{array}{cccc}\sin 2\phi& -(1-\cos 2\phi) \\
	1-\cos 2\phi&\sin 2\phi \end{array}\right).
	\end{split}
	\end{equation}
	Here the abbreviations are $u^{x}=u^{x}(\textbf{r})$, $u'^{x}=u^{x}(R^{-1}\textbf{r})$, and so as $u'^{y}$. Subsequently, one can get the analytic form of $(\partial_{x_{1}}\partial_{x_{2}}-\partial_{x_{2}}\partial_{x_{1}})u'^{x}$, and so as $u'^{y}$.
	
	\section{Interlayer coupling and imaginary Fermi velocity}
	\label{app:appc}
	\begin{figure*}[htp!]
		\includegraphics[width=2\columnwidth]{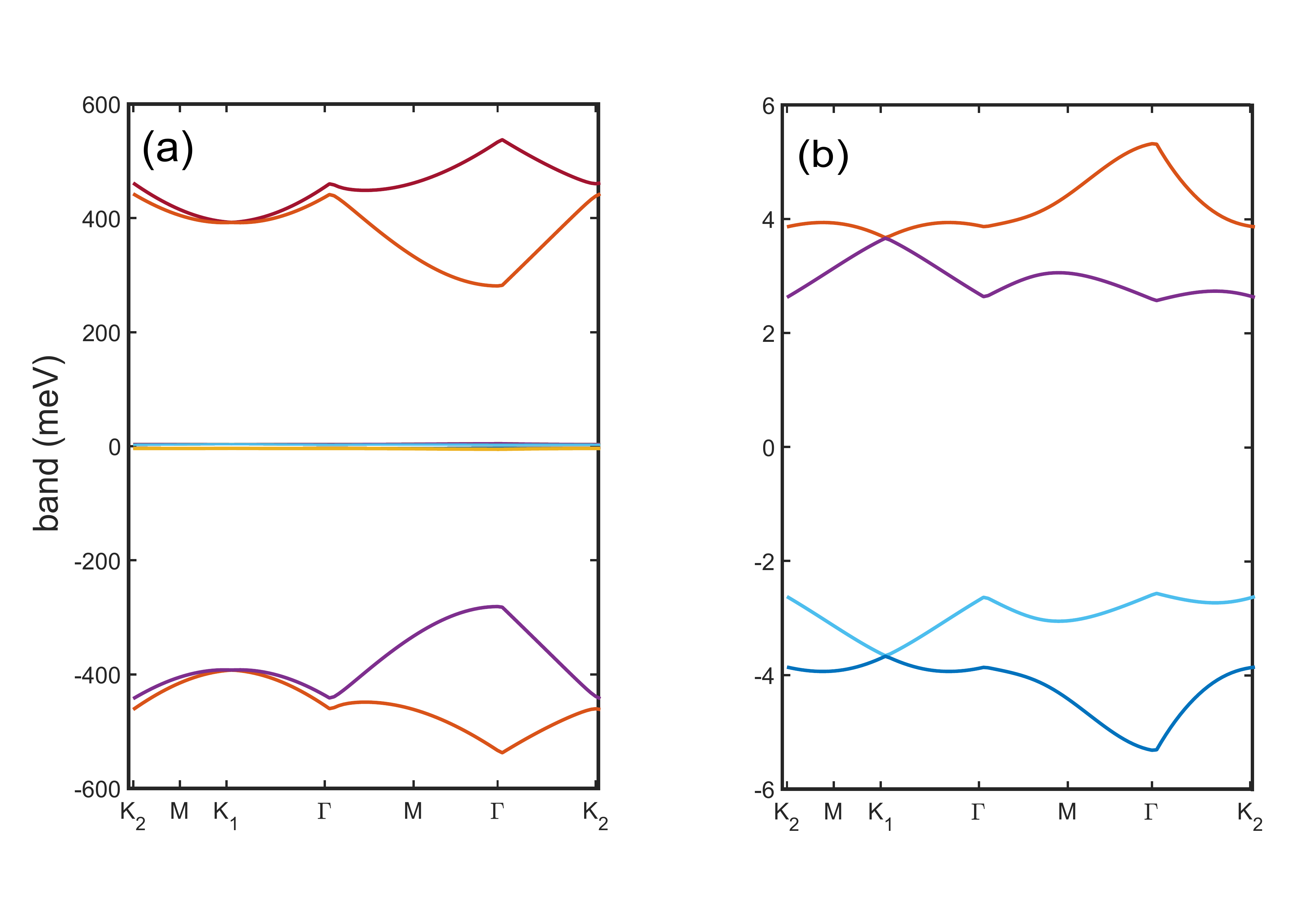}
		\caption{The band structure for $\phi=1.01^{0}$ under imaginary vertical fermi velocity model with $\Gamma=10^{-3}$. The left panel shows 12 bands near $\epsilon=0$ and the right panel shows 8 bands. The lowest bands are even flatter than the counterpart in Fig.~\ref{fig:band}. However the Dirac cone located at $\bf{K}_{1}$ and $\epsilon=0$ vanishes. Other characters are qualitatively the same as the bands in Fig.~\ref{fig:band}.}
		\label{fig:imaginaryvertical}
	\end{figure*}
	Starting from the tight-binding (microscopic) model and considering interlayer coupling as a tunneling interaction, the bare interlayer hopping term can be written as
	\begin{eqnarray}
	&t_{\perp}(c_{1}^{\dagger}c_{2}+c_{2}^{\dagger}c_{1})=t_{\perp}(c_{1}^{\dagger}c_{1+dz}+c_{2}^{\dagger}c_{2-dz})\approx \nonumber \\
	&t_{\perp}(c_{1}^{\dagger}c_{1}+c_{2}^{\dagger}c_{2}+hc_{1}^{\dagger}\partial_{z}c_{1}-hc_{2}^{\dagger}\partial_{z}c_{2})\nonumber \\
	&=t_{\perp}(c_{1}^{\dagger}c_{1}+c_{2}^{\dagger}c_{2})-i[(it_{\perp}h)c_{1}^{\dagger}\partial_{z}c_{1}+(-it_{\perp}h)c_{2}^{\dagger}\partial_{z}c_{2}],
	\end{eqnarray}
	where compared to the mean free path of a Dirac fermion, we have assumed a small interlayer distance. In the above expression $\partial_{z}\psi\approx\frac{1}{h}(\psi(2)-\psi(1))$ where the labels $1,2$ stands for the bottom and top layer, respectively. The vertical Fermi velocity for two layers conjugate to each other is given by $v_{\perp}=\pm i t_{\perp}h$. One can estimate the vertical Fermi velocity by using the Koster-Slater parametrization~\cite{Liang-Fu2018prx}. That is, $\left|\frac{v_{\perp}}{v_{f}}\right|=\left|\frac{t_{\perp}h}{t_{//}a}\right|=\left|\frac{V^{0}_{pp\sigma}h}{V^{0}_{pp\pi}a}\right|=\frac{0.48\times 3.35}{2.7\times 1.42}\approx 0.419$.
	
	The occurrence of imaginary vertical velocity is not unusual (within the context of our calculation) since in the confined direction, the wave function is always decaying. In Fig.~\ref{fig:imaginaryvertical}, we illustrate the lowest bands for curved space-time Dirac model Eq.~\eqref{eq:action} with imaginary vertical Fermi velocity. The parameters are $\phi=1.01^{0}$ and $\Gamma=10^{-3}$. The other parameters are the same as in Fig.~\ref{fig:band}. In the imaginary vertical velocity model, the lowest bands are even flatter than their counterpart in the real velocity model, see Fig.~\ref{fig:band}.
	However the original Dirac cone located at $\bf{K}_{1}$ and $\epsilon=0$ vanishes. Other characteristics are qualitatively the same as the bands in Fig.~\ref{fig:band}. So we conjecture that using the original real velocity model is appropriate. The imaginary vertical velocity model is only used in Fig.~\ref{fig:imaginaryvertical}. We implemented the original real vertical velocity model in the manuscript unless explicitly stated otherwise.
	\bibliographystyle{apsrev4-2}
	\bibliography{RBGmain}
\end{document}